\documentclass[fleqn,usenatbib]{mnras}

\usepackage{newtxtext,newtxmath}

\usepackage[T1]{fontenc}

\DeclareRobustCommand{\VAN}[3]{#2}
\let\VANthebibliography\thebibliography
\def\thebibliography{\DeclareRobustCommand{\VAN}[3]{##3}\VANthebibliography}

\usepackage{graphicx}	
\usepackage{amsmath}	
\usepackage{amssymb}	

\newcommand{\vel}{\mathrm{v}}
\newcommand{\dvrot}{\Delta \left( \delta \vel_{rot} \right)}

\title[Constraining Disk Structure]{Constraining Protoplanetary Disk Accretion and Young Planets Using ALMA Kinematic Observations}

\author[I. Rabago and Z. Zhu]{
Ian Rabago,$^{1}$\thanks{E-mail: rabagoi@unlv.nevada.edu}
Zhaohuan Zhu$^{1}$
\\
$^{1}$Department of Physics and Astronomy, University of Nevada, Las Vegas 4505 S. Maryland Parkway Las Vegas, NV 89154, USA\\
}

\date{Accepted XXX. Received YYY; in original form ZZZ}

\pubyear{2020}

\begin{document}
\label{firstpage}
\pagerange{\pageref{firstpage}--\pageref{lastpage}}
\maketitle

\begin{abstract}
Recent ALMA molecular line observations have revealed 3-D gas velocity structure in protoplanetary disks, shedding light on mechanisms of disk accretion and structure formation. 1) By carrying out viscous simulations, we confirm that the disk's velocity structure differs dramatically using vertical stress profiles from different accretion mechanisms. Thus, kinematic observations tracing flows at different disk heights can potentially distinguish different accretion mechanisms. On the other hand, the disk surface density evolution is mostly determined by the vertically integrated stress. The sharp disk outer edge constrained by recent kinematic observations can be caused by a radially varying $\alpha$ in the disk. 2) We also study kinematic signatures of a young planet by carrying out 3-D planet-disk simulations. The relationship between the planet mass and the ``kink'' velocity is derived, { showing a linear relationship with little dependence on disk viscosity, but some dependence on disk height when the planet is massive (e.g. $10 M_J$). We predict the “kink” velocities for the potential planets in DSHARP disks. }At the gap edge, the azimuthally-averaged velocities at different disk heights deviate from the Keplerian velocity at similar amplitudes, and its relationship with the planet mass is consistent with that in 2-D simulations. After removing the planet, the azimuthally-averaged velocity barely changes within the viscous timescale, and thus the azimuthally-averaged velocity structure at the gap edge is due to the gap itself and not directly caused to the planet. Combining both axisymmetric kinematic observations and the residual “kink” velocity is needed to probe young planets in protoplanetary disks.
\end{abstract}




\section{Introduction}
\label{sec:intro}
Protoplanetary disks are thought to be the sites of planetary formation.  In recent years, use of ground-based radio telescopes such as the Atacama Large Millimeter Array (ALMA) has improved our ability to resolve protoplanetary disks, revealing a wealth of structure in many existing disks, including rings, spirals, and other non-axisymmetric structures (e.g. \citealt{ALMA2015, Andrews2016, Andrews2018}).  It is often suggested that the dark annular features observed in the dust continuum are gaps created by protoplanets as they repel material away from the planet (e.g. review by \citealt{Espaillat2014,Andrews2020}).  Despite the high frequency of gaps observed in high-resolution imagery, only a small number of forming planets have been discovered to date.  One notable protoplanetary system is the disk of PDS 70, which contains two planetary-mass point sources detected by direct observation \citep{Keppler2018, Wagner2018, Haffert2019, Christiaens2019, Isella2019, Wang2020}.  Many other disks have been suspected to contain protoplanets due to the presence of gaps and spirals within the disk, but there is much debate over their existence as these features are not a unique signature of planetary formation (e.g. \citealt{Follette2017,Rameau2017,Cugno2019, Brittain2019}).

Besides planets, there are many other mechanisms that are able to form gaps in protoplanetary disks.  Gaps and other axisymmetric structures can be formed by condensation of materials at snow lines \citep{Zhang2015, Okuzumi2016}, interaction of the disk material with magnetic fields \citep{KretkeLin2007, Johansen2009, Suriano2017,Bai2014, Hu2019, Hu2020}, dust-gas secular gravitational instability \citep{Takahashi2014,Tominaga2020}, and indirectly from the spiral arm generated by a planet orbiting at a different radius \citep{Bae2017, Dong2017}.  Determining which of the observed gaps are hosts to potential planets is an ongoing process and an area of active research.

Thanks to ALMA's high sensitivity to molecular lines,
sub-thermal kinematic motion in protoplanetary disks can be probed using molecular lines, and such kinematic information provides more direct evidence of the planets in disks \citep{Teague2018, Pinte2018}. To extract the disk kinematic information from molecular lines, two different methods have been developed. \cite{Pinte2018} uses the high spectral-resolution channel maps to probe the disk region whose velocity structure is significantly disturbed by the planet. Such a non-axisymmetric disturbance is due to the horseshoe and circumplanetary motion around the planet \citep{Perez2015, Perez2018}, forming a ``kink'' in the channel maps. The second method, used by   
\cite{Teague2018}, assumes that the disk flow is mostly axisymmetric, allowing the axisymmetric flow velocity to be calculated by averaging the kinematic information along the disk's azimuthal direction. With this averaging process, sub-thermal axisymmetric deviations from the Keplerian motion can be extracted.  
\cite{Teague2018} analyzed the rotational velocity structure in the gaps of disk HD 163296 and found that the super-Keplerian and sub-Keplerian rotational velocity at the gap edges could be explained by planets of $1M_J$ at 100 AU and $1.3 M_J$ at 165 AU.  A later work \citep{Teague2019} examines the two-dimensional velocity components of HD 163296, and finds evidence of meridional flows in the atmosphere above the gaps.  Such meridional flow also resembles the flow structure in 3-D planet-disk interaction simulations \citep{FungChiang2016}.

 The relationship between the planet and the ``kink'' velocity has not been derived. \cite{Pinte2018} and \cite{Pinte2019}  derived the planet mass by comparing the observed channel maps with synthetic channel maps from direct numerical simulations. Although such an approach provides a more robust estimate on the planet mass, it is time consuming and computationally expensive for a large sample of disks (e.g. \citealt{Pinte2020}). 

The relationship between the axisymmetric deviation from the Keplerian motion at the gap edge and the embedded planet mass has been 
derived in \cite{Zhang2018} and \cite{GyeolYun2019} using a grid of $R-\phi$ 2-D simulations. 
However, strong molecular lines are normally produced at the disk surface which cannot be studied using $R-\phi$ 2-D simulations. 
Thus, it is unclear if such derived relationships can be applied to real molecular line observations.

Besides probing the planet, the disk's axisymmetric deviation from the Keplerian velocity has also been used to study gaseous rings \citep{Rosotti2020} and the disk's outer edge \citep{Dullemond2020}. Assuming that such velocity deviation is due to the radial pressure gradient in the disk, \cite{Dullemond2020} find that the disk around HD 163296 has a rather abrupt outer edge with $\Sigma\propto exp(-(R/R_d)^2)$. \cite{Dullemond2020} suggest that  it is due to outside-in photoevaporation or truncation 
by an unseen companion. On the other hand, the sharp outer edge could be due to the disk evolution itself, as we will argue in \S 4.1.



In this paper, we use 3-D viscous simulations to study the relationship between the planet mass and the disk's 3-D velocity structure (both the ``kink'' velocity and the axisymmetric deviation from the Keplerain velocity at the gap edge). We also
explore how probing the velocity structure could differentiate different accretion processes in protoplanetary disks. The disk could have complicated flow structure (e.g. meridional flow) even without gaps \citep{Urpin1984, KleyLin1992,Pozyczka1994}. We will address how such flow pattern is related to the stress distribution in the disk, and how the disk surface density evolution depends on the stress distribution. Finally, we study the differences between the velocity structure of a gap which has a planet in it and that of a gap without a planet. In Section \ref{sec:visc} we outline the viscous theory important to this work.  We describe our simulation setup in Section \ref{sec:method}, and present our results in Section \ref{sec:results}.  We discuss important results and implications of our findings in Section \ref{sec:discussion}.  Finally, we conclude in Section \ref{sec:conclusion}.



\section{Viscous Theory} \label{sec:visc}
{ Before we carry out viscous simulations for planet-disk interaction, we will first study the evolution of viscous disks without planets for two reasons. 
First, there is a well developed analytical theory on viscous disk structure/evolution, which can be used to test the viscosity module for our numerical simulations. 
Second, by modifying the traditional viscous disk theory, we can derive the flow structure of disks undergoing different instabilities, and how kinematic observations may be able
to constrain these different accretion mechanisms.}

In this section, a few analytical results from the viscous disk theory are reviewed, starting from one-dimensional surface density evolution to two-dimensional meridional circulation. Various stress profiles have been adopted to mimic the turbulent structure in disks subject to different disk instabilities. Then, motivated by some recent MHD simulations showing that the disk mostly accretes in the radial direction under the spherical-polar coordinate system, we analytically derive the meridional circulation pattern under the spherical-polar coordinate system. This new analytical derivation can also be compare with our spherical-polar simulations more directly in Section \ref{sec:method}.

\subsection{1-D Disk Evolution}
\label{sec:1d}
If the viscous disk has an internal viscosity of $\nu$, the disk's surface density ($\Sigma$) follows the diffusion equation
\begin{equation}
    \frac{\partial \Sigma}{\partial t} = \frac{3}{R} \frac{\partial}{\partial R} \left[ R^\frac{1}{2}  \frac{\partial}{\partial R}\left( R^\frac{1}{2}\nu \Sigma \right) \right]\,.
\end{equation}
The analytical solution of this equation has been derived in \cite{LyndenBellPringle1974}.  For a Keplerian disk with 
\begin{equation}
\nu \propto R^\gamma\,,\label{eq:nugamma}
\end{equation}
one similarity solution is (we adopt the derivation from \citealt{Hartmann1998a})
\begin{equation}
    \Sigma(r,t) = \frac{C}{3 \pi \nu} T^{-\frac{5/2-\gamma}{2-\gamma}} exp \left( \frac{-\left(R/R_1\right)^{2-\gamma}}{T} \right)\label{eq:surfevo}
\end{equation}
\begin{equation}
    \dot{M}(r,t) = C T^{-\frac{5/2-\gamma}{2-\gamma}} exp \left( \frac{-\left(R/R_1\right)^{2-\gamma}}{T} \right) \times \left[ 1 - \frac{2(2-\gamma)\left(R/R_1\right)^{2-\gamma}}{T} \right]
\end{equation}
with $T = t/t_s + 1$,
where the viscous timescale is
\begin{equation}
    \label{eq:viscts}
    t_s = \frac{1}{3(2-\gamma)^2}\frac{R_1^2}{\nu_1}\,,
\end{equation}
where $\nu_1=\nu(R_1)$.
The scaling factors for the surface density, radius, and time in this similarity solution are $C$, $R_1$, and $t_s$ respectively. 
Note that the surface density profile at one specific time follows
\begin{equation}
    \label{eq:sdprof}
    \Sigma(R) \propto \nu^{-1} exp \left( -\left(R/R_2\right)^{2-\gamma} \right) \propto R^{-\gamma} exp \left( -\left(R/R_2\right)^{2-\gamma} \right)
\end{equation}
where $R_2 = R_1 T^{1/(2-\gamma)}$. The widely used relationship $\Sigma \propto R^{-1} exp(-R/R_2)$ (e.g. \citealt{Hartmann1998b}; \citealt{Andrews2009}) is the special case with $\gamma=1$.
Considering that $\nu=\alpha c_s^2/\Omega$, $\gamma=1$ corresponds to a constant $\alpha$ value in a disk having $T \propto R^{-1/2}$ (close to the temperature of a passively irradiated disk). 

\subsection{\texorpdfstring{$R-z$ 2-D Meridional Circulation}{R-z 2-D Meridional Circulation}}
\label{sec:meridian}
Although the disk's 1-D evolution is straightforward, the disk's 2-D ($R-z$) accretion structure is much more complicated.
The disk's $R-z$ 2-D flow structure needs to be self-consistently 
solved using the fluid equations in the $R-z$ plane. The disk is unlikely to accrete at the same speed at different heights if the disk's vertical structure is considered. The resulting flow pattern shows the meridional circulation. 
Assuming that the initial density profile at the disk midplane is
\begin{equation}
\rho_{0}(R,z=0)=\rho_{0}(R_{0},z=0)\left(\frac{R}{R_{0}}\right)^p\,,\label{eq:rho}
\end{equation}
and the temperature varies radially (but constant on cylinders)
\begin{equation}
T(R,z)=T(R_{0})\left(\frac{R}{R_{0}}\right)^q\,,
\end{equation}
the hydrostatic equilibrium in the $R-z$ plane requires that (e.g. \citealt{Nelson2013})
\begin{equation}
\rho_{0}(R,z)=\rho_{0}(R,z=0) {\rm exp}\left[\frac{GM}{c_{s}^2}\left(\frac{1}{\sqrt{R^2+z^2}}-\frac{1}{R}\right)\right]\,,\label{eq:rho0}
\end{equation}
and
\begin{equation}
\vel_{\phi}(R,z)=\vel_{K}\left[(p+q)\left(\frac{c_{s}}{\vel_{K}}\right)^2+1+q-\frac{qR}{\sqrt{R^2+z^2}}\right]^{1/2}\,,\label{eq:vphi}
\end{equation}
where $c_{s}=\sqrt{p/\rho}$ is the isothermal sound speed, $\vel_{K}=\Omega_{K}R=\sqrt{GM_{*}/R}$, and $H=c_{s}/\Omega_{K}$.  

The disk's radial velocity under the cylindrical coordinate system can be derived using the angular momentum equation
\begin{equation}
\frac{\partial \rho\delta \vel_{\phi}}{\partial t}=-\frac{1}{R^2}\frac{\partial R^2T_{R\phi}}{\partial R}-\frac{\rho \vel_{R}}{R}\frac{\partial R^2 \Omega_{K}}{\partial R}-\frac{\partial T_{\phi z}}{\partial z}-\rho \vel_{z}\frac{\partial R\Omega_{K}}{\partial z}\,,
\end{equation}
where $\delta \vel_{\phi}=\vel_{\phi}-R\Omega_{K}$, and $T$ can represent the turbulent Reynolds stress, Maxwell stress, or/and viscous stress. 
This equation applies to not only axisymmetric flows but also  non-axisymmetric flows simply by replacing every term with the azimuthally averaged quantities.

For steady flows (constant $\rho\delta \vel_{\phi}$) or axisymmetric flows ($\delta \vel_{\phi}$=0),  we can derive
\begin{equation}
   \frac{\rho \vel_{R}}{R}\frac{\partial R^2 \Omega_{K}}{\partial R}=-\frac{1}{R^2}\frac{\partial R^2T_{R\phi}}{\partial R}-\frac{\partial T_{\phi z}}{\partial z}\,.
\end{equation}
after neglecting higher order terms ($(H/R)^2$ and above).  
For viscous fluid having $T_{R\phi}=\mu_{R\phi} R\partial \Omega/\partial R$ and $T_{\phi z}=\mu_{\phi z}\partial \vel_{\phi}/\partial z$, if the viscosity is isotropic with $\mu_{R\phi}=\mu_{\phi z}=\mu$, we have
\begin{equation}
\frac{\rho \vel_{R}}{R}\frac{\partial R^2 \Omega_{K}}{\partial R}=-\frac{1}{R^2}\frac{\partial \left(R^3\mu\frac{\partial \Omega}{\partial R}\right)}{\partial R}-\frac{\partial\left(\mu\frac{\partial \vel_{\phi}}{\partial z}\right)}{\partial z}\,.\label{eq:fullapha}
\end{equation}
If we assume $\mu=\rho\nu$, $\nu=\alpha c^2_{s}/\Omega$, and $\alpha=\alpha_0 (R/R_0)^{s}$, and then plug in the density and velocity profiles, we can derive
\begin{equation}
\vel_{R}=-\frac{\nu}{R}\left[3s+3p+2q+6+\frac{5q+9}{2}\left(\frac{z}{H}\right)^2\right]\,.\label{eq:vR1}
\end{equation}
as in \cite{Urpin1984,KleyLin1992,Pozyczka1994,TakeuchiLin2002}.
Thus, if $3s+3p+2q+6<0$, the flow is outwards at the disk midplane. For a commonly adopted disk structure with $s=0$, $p=-2.25$, and $q=-0.5$, the quantity $3s+3p+2q+6=-1.75$ so that the flow is outwards at the midplane. Since $5q+9$ is larger than 0 in this case, the flow can be inwards at the disk surface when $z/H$ is large enough.

On the other hand, turbulence in protoplanetary disks may be highly anisotropic. For example, local MHD shearing box simulations suggest that turbulence induced by the magneto-rotational instability (MRI) generates much stronger $R-\phi$ stress than the $\phi-z$ stress \citep{HGB1995}. However, hydrodynamical simulations suggest that the $\phi-z$ stress is much stronger than the $R-\phi$ stress for the vertical shear instability (VSI) \citep{Nelson2013,Stoll2017,Flock2020}.

If we can ignore the $\phi-z$ stress (e.g. for MRI generated turbulence), we can derive
\begin{equation}
\vel_{R}=-\frac{\nu}{R}\left[3s+3p+3q+6+\frac{3q+9}{2}\left(\frac{z}{H}\right)^2\right]\,.\label{eq:vR2}
\end{equation}
as in \cite{Fromang2011, Jacquet2013,PhilippovRafikov2017}. For cases with $\mu_{\phi z}$ significantly larger than $\mu_{R\phi}$ (e.g. VSI), we can assume $\mu_{\phi z}=C\mu_{R\phi}$ (which is equivalent to $\nu_{\phi z}=C\nu_{R\phi}$ or $\alpha_{\phi z}=C\alpha_{R\phi}$) and we can derive
\begin{equation}
\vel_{R}=-\frac{\nu_{R\phi}}{R}\left[3s+3p+3q+6+\frac{3q+9}{2}\left(\frac{z}{H}\right)^2-C q\left( 1- \left(\frac{z}{H}\right)^2\right)\right]\,.\label{eq:vR3}
\end{equation}

{ where $\nu_{R\phi} = \mu_{R\phi}/\rho$ is the viscosity generated from the $R-\phi$ stress.}

At the disk midplane, this equation recovers Equation 12 in \cite{Stoll2017}. When $3s+3p+3q+6-Cq<0$, the flow is inwards at the midplane. With $q=-0.5$ and $C=650$  \citep{Stoll2017} for VSI due to the significant motion in the vertical direction \citep{LinYoudin2015}, the disk flows inwards at the midplane and outwards at $z \gtrsim H$.

However, all above derivations are based on the $R-z$ cylindrical coordinate system. Some instabilities (e.g. VSI) show clear distinction between $R-\phi$ and $\phi-z$ stresses, so that a cylindrical coordinate system is more appropriate to study disk accretion in these disks \citep{Stoll2017}. On the other hand, some other disks (e.g. MHD disks) have accretion flow  which is in a direction more aligned with the radial direction in the spherical-polar coordinate system (e.g. \citealt{ZhuStone2018}), and the $r-\phi$ stress and $\theta-\phi$ stress play more distinct roles on disk accretion. 
This motivates us to derive the meridional circulation under the spherical-polar coordinate system, which  also makes it easier to compare with our spherical-polar numerical simulations in Section \ref{sec:method}. The angular momentum equation under the spherical-polar coordinate system is
\begin{align}
\frac{\partial \rho\delta \vel_{\phi}}{\partial t}&=-\frac{1}{r^3}\frac{\partial r^3   T_{r\phi}}{\partial r}-\frac{\rho \vel_{r}}{r}\frac{\partial r^2\Omega_{K}}{\partial r}\nonumber\\
&-\frac{1}{r {\rm sin}^2\theta}\frac{\partial {\rm sin}^2\theta T_{\theta\phi}}{\partial \theta}-
\frac{\rho \vel_{\theta}}{r {\rm sin}\theta}\frac{\partial{\rm sin}\theta r \Omega_{K}}{\partial \theta}\label{eq:angsph}\,,
\end{align}
where $\vel_{K}=r\Omega_{K}$. Throughout the paper, to distinguish the radial direction between the cylindrical and spherical-polar coordinate systems, we use $R$ to represent the radial direction in the cylindrical coordinate system, and $r$ for the radial direction in the spherical-polar coordinate system. For steady flows with second and higher order terms removed, we have
\begin{equation}
    \frac{\rho \vel_{r}}{r}\frac{\partial r^2\Omega_{K}}{\partial r}=-\frac{1}{r^3}\frac{\partial r^3   T_{r\phi}}{\partial r}
-\frac{1}{r {\rm sin}^2\theta}\frac{\partial {\rm sin}^2\theta T_{\theta\phi}}{\partial \theta}\,.
\end{equation}

Although solving the same fluid equations under different coordinate systems won't change the results, $T_{\theta\phi}$ and $T_{\phi z}$ are very different at the disk surface due to the different curvatures of these two systems, and ignoring $T_{\theta\phi}$ will lead to a different flow structure than Equation \ref{eq:vR2} which ignores $T_{\phi z}$.
If we ignore the $\theta$-$\phi$ stress and use $T_{r\phi}=\mu r \partial \Omega_{K}/\partial r$,
the disk's radial velocity is
\begin{equation}
\frac{\rho \vel_{r}}{r}\frac{\partial r^2 
\Omega_{K}}{\partial r}=-\frac{1}{r^3}\frac{\partial \left(r^4\mu\frac{\partial \Omega_K}{\partial r}\right)}{\partial r}\,.\label{eq:vre}
\end{equation}
Assuming $\mu=\rho\nu$ and $\alpha=\alpha_0 (r/r_0)^{s}$ in the spherical-polar coordinate system, we can derive
\begin{equation}
\vel_{r}=-\frac{\nu}{r}\left[3s+3p+3q+9+\frac{3q+3}{2}\left(\frac{z}{H}\right)^2\right]\,.\label{eq:vr2}
\end{equation} 
By comparing Equation \ref{eq:vR2} with Equation \ref{eq:vr2}, we can see that ignoring $T_{\theta\phi}$ or $T_{\phi z}$ leads to different flow structures. It is important to choose the appropriate coordinate system so that the different components of the stress from turbulence can be more clearly separated (e.g. some instabilities may generate turbulence with $T_{\theta\phi}$=0 but non-zero $T_{\phi z}$.).

One particular stress profile that is motivated by MHD simulations is that the  stress is vertically uniform until a certain number of disk scale heights (z=$h_{cut}$H) \citep{Fromang2011}. Thus, we write $\mu$ in Equation \ref{eq:vre} as
\begin{align}
\mu=\left \{
  \begin{tabular}{ll}
  $\alpha \rho_0\frac{c_{s}^2}{\Omega_K}\frac{\sqrt{2\pi}}{2 h_{cut}}$ & $|z| \leq h_{cut}H$ \\ \\
  0 & |z| > $h_{cut}H$
  \end{tabular}
  \right.\label{eq:unistress}
\end{align}
so that the vertically integrated stress is still
$\int\mu dz=\alpha \Sigma c_{s}^2/\Omega$.
If $\mu$ can be written as $\mu_0(r/r_0)^\eta$ within $h_{cut}H$, the radial velocity within $h_{cut}H$ is
\begin{equation}
    \vel_{r}=-(4.5+3\eta)\mu_0 \left(\frac{r}{r_0}\right)^{\eta}\frac{1}{r \rho }\,.\label{eq:vr3}
\end{equation}
where $\eta=s+p+q+1.5$.
Comparing Equation \ref{eq:vr3} with Equation  \ref{eq:vr2}, we can see that the uniform stress case has a sharp increase of velocity at the disk surface (exponential increase due to the 1/$\rho$ dependence). 

Different assumptions on the vertical stress profiles due to different accretion mechanisms lead to dramatically different meridional flow structures.
The analytical solutions for the meridional circulation (Equations \ref{eq:vR1}, \ref{eq:vR2}, \ref{eq:vR3}, \ref{eq:vr2}, \ref{eq:vr3}) will be compared with our direct numerical simulations in Section \ref{sec:method} to mutually verify the analytical derivation and numerical simulations. 



\section{Methods}
\label{sec:method}

We solve the compressible Navier-Stokes equations using the grid-based magnetohydrodynamics (MHD) code ATHENA++ \citep{Stone2020}.  We adopt the spherical-polar coordinate system $(r, \theta, \phi)$ for the simulations.

\subsection{2-D disk simulations without planets}
\begin{figure}
 \includegraphics[trim=0 0cm 0 0,clip,width=\columnwidth]{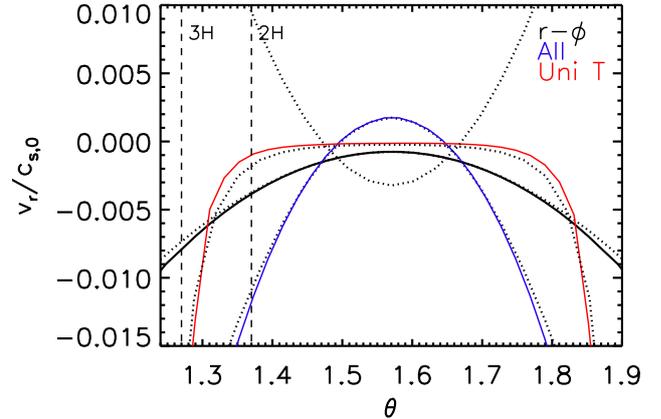}
 \caption{Comparison of the disk radial velocity profiles at r=1 in direct numerical simulations with $\alpha=0.01$ (solid curves) to the analytic equations (dotted curves). The disk achieves this profile at a time of 30 orbits at r=1. The three dotted curves which have negative $\vel_r$ at the disk surface are analytical solution from Equation \ref{eq:vR1}, \ref{eq:vr2}, and \ref{eq:vr3} with $\alpha_0=0.01$. The dotted curve with the positive $\vel_r$ at the surface is from Equation \ref{eq:vR3} with $\alpha_0=10^{-4}$ and C=650 representing the stress derived in the vertical shear instability simulations. The velocity structure is significantly affected by different vertical stress profiles, although all the $\alpha_0=0.01$ cases have the same vertically integrated $r-\phi$ stress. }
 \label{fig:codetest}
\end{figure}

We first carry out $r-\theta$ 2-D simulations without planets to verify our code by comparing the simulation results with Equations \ref{eq:vR1}, \ref{eq:vr2}, and \ref{eq:vr3}. We set up the disk in the domain of $r=[0.3, 3.0]$, and $\theta=[\pi/2-0.5, \pi/2+0.5]$. In the $r$ direction, we have 112 grid cells that are logarithmically uniformly spaced. In the $\theta$ direction, we have 48 uniformly spaced cells. The disk setup follows Equations \ref{eq:rho} to \ref{eq:vphi} with $H/R=0.1$ at R=1, $p = -2.25$ and $q = -0.5$.  We have adopted the reflecting boundary condition in the $\theta$ direction and fixed the inner boundary to the initial condition throughout the simulation. The disk relaxes to the initial temperature at a cooling time of $t_c = 0.01$ orbital time ($2\pi/\Omega$, \citealt{Zhu2015}). Such cooling time corresponds to the radiative cooling timescale at 100 au and is not short enough to trigger VSI \citep{LinYoudin2015}. The $\alpha$ parameter is 0.01. We have carried out three separate simulations with the full viscous stress (Eq. \ref{eq:fullapha}), only the $r-\phi$ stress, and the vertically uniform stress (Eq. \ref{eq:unistress} with $h_{cut}$=4). The radial velocities at $r=1$ and  $t=30\ T_{0}$ are shown in Figure \ref{fig:codetest}, where $T_{0}=2\pi/\Omega_0$ is the orbital time at $r=1$. At 30 orbits, the meridional circulation has been fully established. The three dotted curves which have negative $\vel_r$ at the surface in Figure \ref{fig:codetest} are the analytical solutions from Equations \ref{eq:vR1}, \ref{eq:vr2}, \ref{eq:vr3} with the same parameters \footnote{$z$ and $R$ in these equations are calculated using $r$ and $\theta$ of each cell in simulations.}.  We can see good agreements between the simulations and the analytical solutions. For comparison, the dotted curve with the positive $\vel_r$ at the surface is from Equation \ref{eq:vR3} with $\alpha_0=10^{-4}$ and C=650 representing the stress similar to those derived from the vertical shear instability simulations \citep{Stoll2017}. 

After testing our code against analytical solutions, we carry out a set of simulations to study the disk surface density evolution for disks having these different vertical stress profiles. These simulations have similar setups as the previous tests except for a few modifications. To study the disk's viscous spread, We setup a Gaussian bump at $r=1$ with a Gaussian width of $0.1$. We also allow $\alpha$ to change with radii ($\gamma \neq 1$ in Equation \ref{eq:nugamma}). We have 680 logarithmically uniformly spaced grids in the radial direction for the domain $r=[0.1,20]$. We have 128 uniformly spaced grid cells in the $\theta$ direction in the domain $\theta = [\pi/2-0.5, \pi/2+0.5]$. The outflow boundary condition and reflecting boundary condition are adopted in the radial and $\theta$ direction.  { Our outflow boundary condition, which does not allow the gas to flow from the ghost zones into the computational domain, is different from the default outflow boundary condition in Athena++.}  The simulations are run for 200 $T_0$.

\subsection{3-D planet-disk interaction simulations}
After studying the disk surface density evolution under different stress profiles, we carry out 3-D viscous simulations to study the disk's velocity structure influenced by a young planet. The simulation domain covers a range of $r = [0.3, 3.0]$, $\theta = [\pi/2-0.5, \pi/2+0.5]$, and $\phi = [0, 2\pi]$.  The domain is divided into 112 logarithmically spaced cells in $r$, 48 uniformly spaced cells in $\theta$, and 152 uniformly spaced cells in $\phi$.  The number of cells in the $\phi$ direction is half the number required to ensure square cells in the $r-\phi$ plane; many quantities examined in this paper are azimuthally averaged, therefore this reduction in cells is expected to have a small effect.


To initialize the density profile of the disk, we numerically integrate the density at each grid cell to establish vertical hydrostatic equilibrium.  We initialize the sound speed profile using the power-law profile
\begin{equation}
    c_s = c_{s,0} \left( \frac{r}{r_0}\right)^{-q}\,.
\end{equation}
As with the 2-D simulations, we use power-law exponents of $p = -2.25$ and $q = -0.5$ for the density and sound speed profiles, respectively.  This gives the disk a surface density profile of $\Sigma \propto r^{-1}$, consistent with observations of older protoplanetary disks \citep{Andrews2009}.  The disk again has a cooling time of $t_c = 0.01$ orbital time and $H/R = 0.1$ at $r_0$.  We use reflecting boundary conditions in the $\theta$ direction and fix the radial boundary conditions to their initial values throughout the simulation.



The disk velocity is initialized with only the rotational component $\vel_\phi$.  To initialize the radial velocity profile, we allow the disk to evolve for a period of roughly $25\ T_0$ and allow the gas to settle into a steady state.  With no initial velocity in the $r-\theta$ plane, the radial velocity of the disk will settle towards Equation \ref{eq:vR1}.   Once the radial velocity of the disk is properly established, we add a planet at a distance of $r = r_p = 1$ from the central star.  The planet increases in mass over the next 25 orbits from zero up to its final mass $M_p$.  The gravitational potential of the planet is written as a second order potential (e.g. \citealt{Dong2011}):
\begin{equation}
    \Phi_p = -GM_p \frac{1}{\left( r^2 + r_s^2 \right)^{1/2}}\,,
\end{equation}
where $r_s$ is the smoothing radius.  We choose the value of $r_s$ to be 0.1 Hill radii for each simulation.  After reaching its final mass, the planet continues to open a gap in the disk for another 450 planetary orbits, giving the simulation a total time of $500\ T_0$.  

The 3-D simulations have the isotropic disk viscosity $\alpha$ and the final planet mass $M_p$ as free parameters.  We choose final planet masses of $10^{-3}$ and $10^{-2}$ central star mass (which are 1 $M_J$ and 10 $M_J$ if the central star is a solar mass star) in combination with a constant $\alpha$ of $10^{-3}$ and $10^{-2}$, creating a set of four different simulations.  { We also run a second set of simulations to study how the choice of the smoothing length affects the unbound velocity flow in the vicinity of the planet, which we discuss in Section \ref{sec:nonsymvel}.  These simulations use the same disk viscosities of $\alpha = 10^{-3}$ and $10^{-2}$, final planet masses of 1 $M_J$ and 10 $M_J$, but a constant smoothing length of 2 grid cells.}  In order to examine the effect of MHD disks on meridian circulation, we run an additional simulation with a final planet mass of $1M_J$ and a vertically-varying $\alpha$ according to Equation \ref{eq:unistress}, with $h_{cut} = 4$.



\begin{figure}
 \includegraphics[width=\columnwidth]{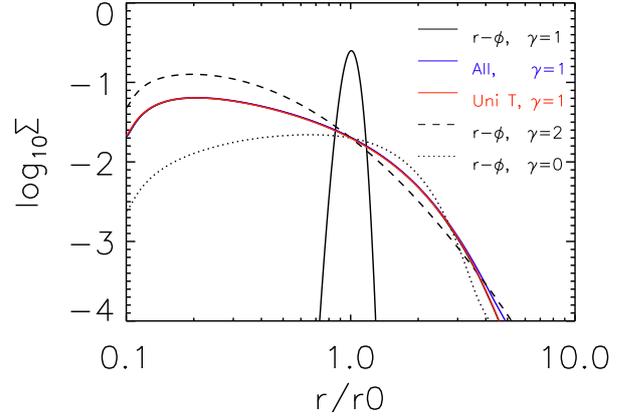}
 \caption{Viscous disk evolution for disks with different stress structures. The Gaussian profile at r=r$_0$ is the initial condition. All other curves are the disk surface density at 200 $T_0$ where $T_0$ is the orbital time at r=r$_0$.}
 \label{fig:example}
\end{figure}



\section{Results}
\label{sec:results}

\subsection{Disk Density Evolution with Different Stress Profiles}
Figure \ref{fig:codetest} shows that the disk's velocity structure is significantly affected by different vertical stress profiles in disks without planets. At two disk scale heights ($\theta\sim1.37$), the full stress case has an inward velocity $\gtrsim$0.01$c_s$, the $r-\phi$ stress only case has an inward velocity of $\sim 0.004c_s$, the uniform stress case has an inward radial velocity $\lesssim$0.001$c_s$,  and the VSI case has an outward velocity $\gtrsim$0.01$c_s$. However, at three disk scale heights, the inward radial velocity in the uniform stress case quickly increases to $\gtrsim$0.015$c_s$. Note that all three cases with $\alpha=0.01$ in  Figure \ref{fig:codetest} have the same vertically integrated $r-\phi$ stress, which means that they have the same $\alpha$ value in the traditional 1-D viscous disk theory. If we also set $\alpha=0.01$ in the VSI stress profile, the velocity profile will be amplified by 100 and out of the plotted range.
Considering that strong molecular lines (e.g. CO) mostly trace the disk atmosphere, the probed velocity will be very sensitive to different vertical stress profiles in the disk, even if different stress profiles correspond to the same $\alpha$ value in 1-D disk evolution. We caution that this traditional viscous stress model has limitations to capture internal stresses due to large scale magnetic fields (e.g. \citealt{ZhuStone2018}) or external stresses due to magnetocentrifugal wind \citep{Bai2016}.
Nevertheless, future kinematic observations using various molecular lines tracing different disk heights will not only measure the value of $\alpha$ but also constrain the detailed vertical stress profiles and accretion mechanisms. 

These largely different velocity structures with different vertical stress profiles raise the question as to whether the surface density evolution of 3-D viscous disks are also affected by the detailed vertical stress profiles, so that the traditional 1-D disk evolutionary model (Equation \ref{eq:surfevo}) cannot capture the disk evolution properly. Thus, we have carried out axisymmetric 2-D simulations with different stress profiles in the disk to study the spread of a Gaussian bump. The results are shown in Figure \ref{fig:example}. With the same vertically integrated stress, all three simulations with three different vertical stress profiles (black, blue, and red solid curves) show exactly the same surface density structure at 200 $T_0$, following a power law $R^{-1}$ before the exponential decrease (Equation \ref{eq:sdprof}). This demonstrates that the detailed vertical stress profiles do not affect the disk surface density evolution, and the traditional 1-D viscous model is sufficient for studying the disk surface density evolution. 

As shown in Figure \ref{fig:example}, we have also carried out full stress simulations whose vertically integrated stress varies along the radial direction steeper or flatter than $\nu\propto R$. 
If the vertically integrated stress changes slower with radii (e.g. $\gamma=0$ in Equation \ref{eq:nugamma} which means $\alpha\propto R^{-1}$ with $T\propto R^{-1/2}$), the surface density follows a flatter power law  before a steeper exponential decrease, as predicted by 1-D model in Equation \ref{eq:sdprof}. 

Recent kinematic measurements from \cite{Dullemond2020} have suggested that, at the disk outer edge of HD 163296, the disk surface density falls faster than $exp(-R/R_d)$ and is closer to $exp(-(R/R_d)^2)$.  Although \cite{Dullemond2020} suggests that this sharper density drop is due to outside-in photoevaporation or truncation by an unseen companion, we suggest that it can also be due to the viscosity's power law having $\gamma < 1$ or $\alpha$ decreases with larger radii.  For example, when $\gamma = 0$ or $\alpha \propto R^{-1}$ with $T\propto R^{-1/2}$, Equation \ref{eq:sdprof} becomes $\Sigma \propto R^0 exp(-(R/R_2)^2)$ (the dotted curve in Figure \ref{fig:example}).  This sharper drop is then consistent with observations in \cite{Dullemond2020}. Another implication of the decreasing $\alpha$ with the increasing radius is that the density profile at the inner disk would be a constant with radii ($\propto R^0$). Some disks indeed show a relatively flat surface density profile (e.g. \citealt{Carrasco2019}). On the other hand, measurements of the disk surface density at the inner disk could be very uncertain due to the optical depth effects \citep{Liu2019,Zhu2019,Carrasco2019}.

\subsection{Non-axisymmetric Velocity Structure Induced by the Planet}
\label{sec:nonsymvel}

\begin{figure}
 \includegraphics[width=\columnwidth]{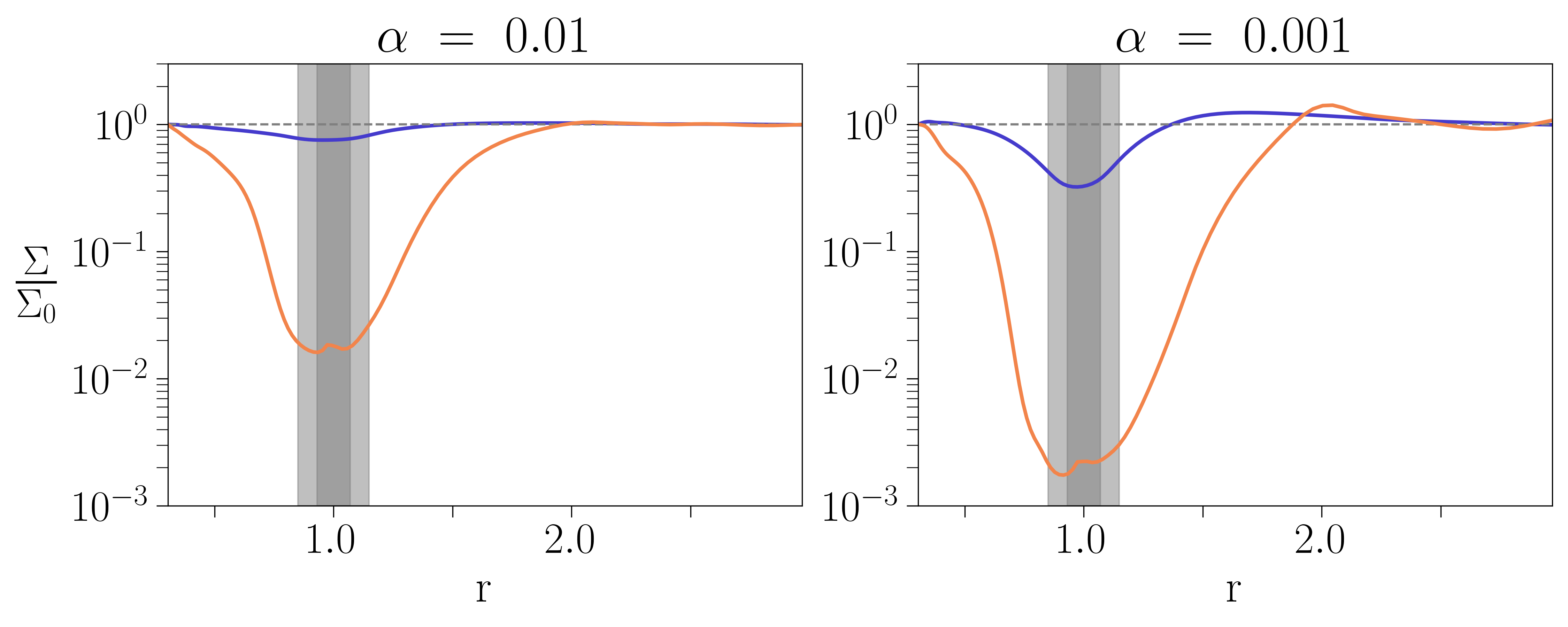}
 \caption{Radial surface density profiles for the $\alpha$=0.01 simulations (left) and $\alpha$ = 0.001 simulations (right) at $t = 500\ T_0$. Blue curves indicate a planet of 1 Jupiter mass, and orange curves indicate a planet of 10 Jupiter masses.}
 \label{fig:surfacedensity}
\end{figure}

\begin{figure*}
 \includegraphics[width=0.95\textwidth]{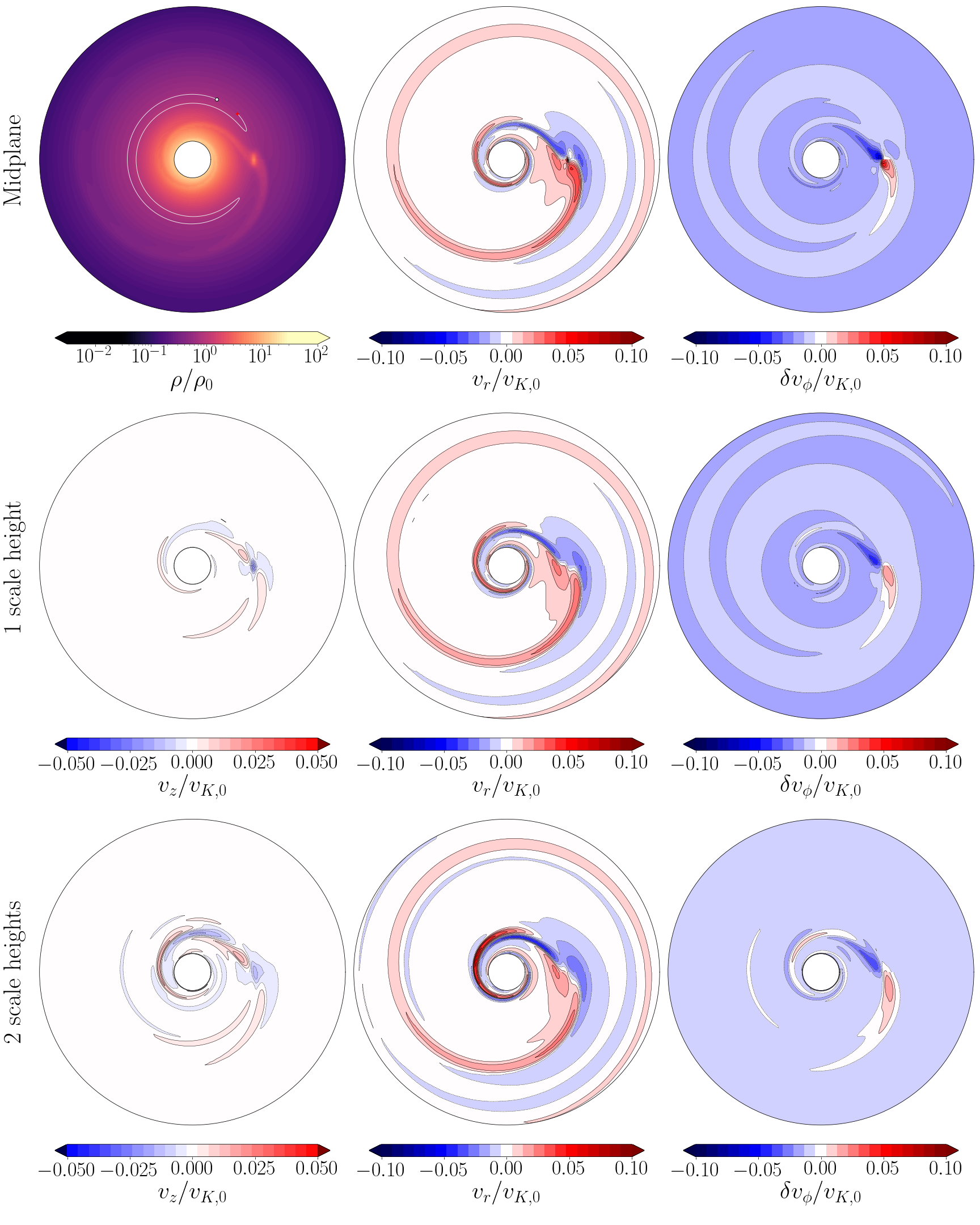}
 \caption{ Midplane cuts of the $\alpha = 0.01,\ M=1M_J$ simulation at $t = 500\ T_0$.  Each row shows the velocity components of the disk at different disk heights (midplane, 1 scale height, and 2 scale heights).  \emph{Top left:} Density plot, with a horseshoe trajectory also plotted within the gap region, beginning at the red marker.  \emph{Left column:} vertical velocity $\vel_z$. \emph{Middle column:} Radial velocity.  The inward/outward radial velocity trace the inner/outer spirals. \emph{Right column:} Deviation from Keplerian velocity.  Regions of sub-Keplerian and super-Keplerian velocity trace the inner and outer gap edges.  All velocities are scaled to the local Keplerian velocity at $R=1$.}
 \label{fig:horseshoe}
\end{figure*}

\begin{figure*}
 \includegraphics[width=0.95\textwidth]{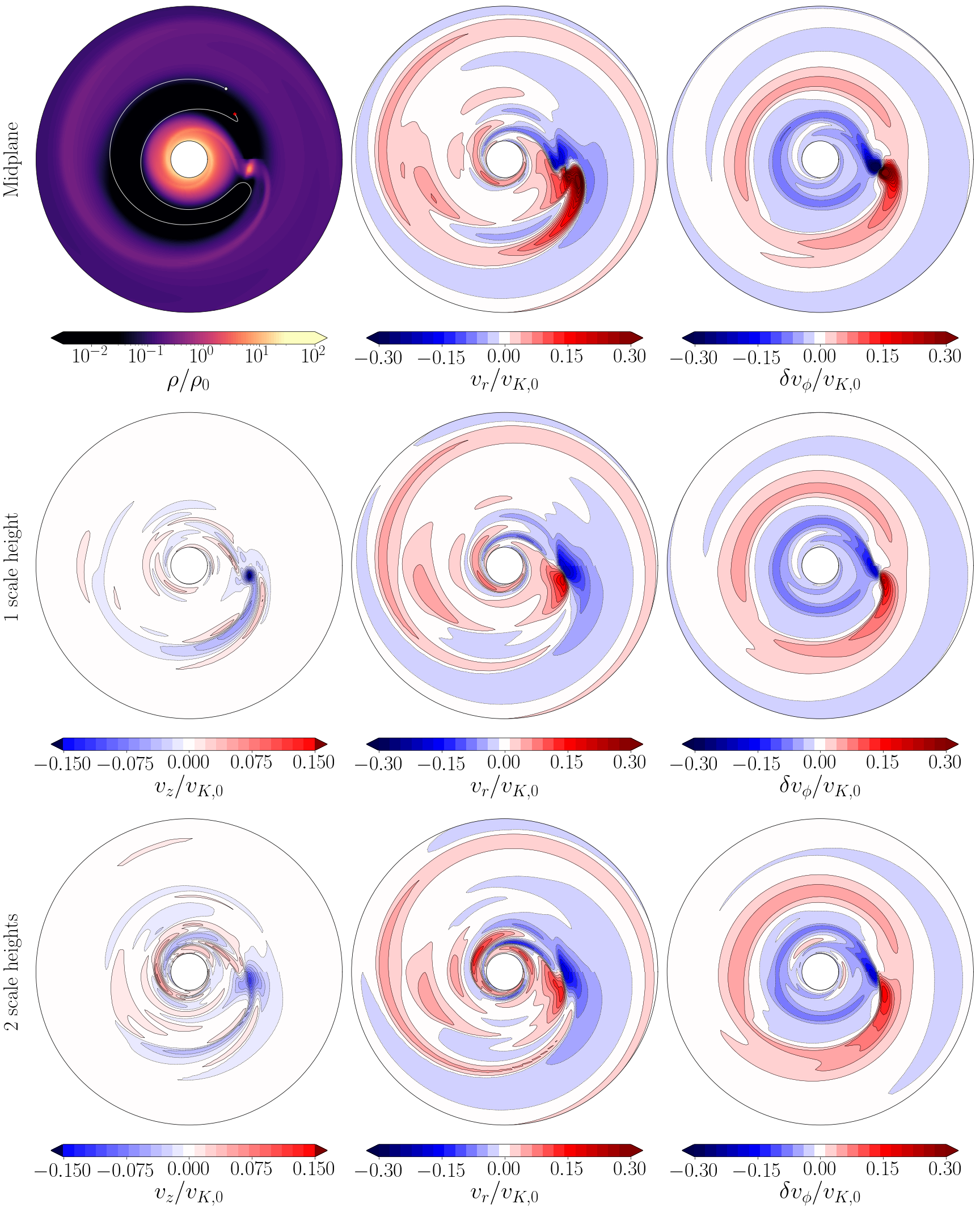}
 \caption{ Same as Figure \ref{fig:horseshoe}, but for the $\alpha = 0.01,\ M=10M_J$ simulation.}
 \label{fig:horseshoe10MJ}
\end{figure*}

\begin{figure*}
 \includegraphics[width=0.98\textwidth]{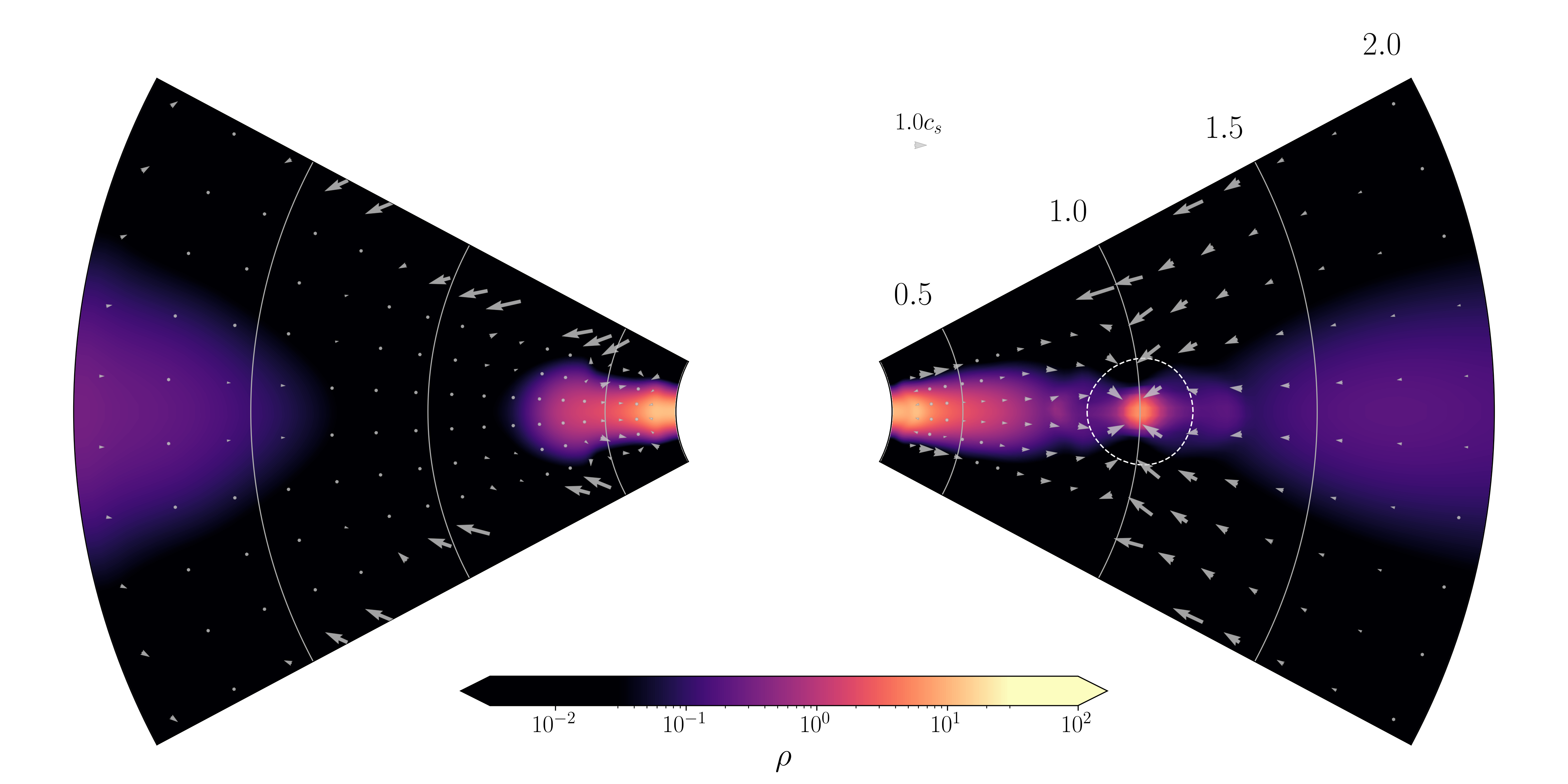}
 \caption{Vertical slice of the $\alpha = 0.01,\ M=10M_J$ simulation at $t = 500\ T_0$.  Background contours show gas density.  Dashed circle on the right wedge denotes the planet's Hill radius.  Vectors denote the gas velocity, scaled to the local sound speed.}
 \label{fig:vslice}
\end{figure*}

\begin{figure}
    \includegraphics[width=\columnwidth]{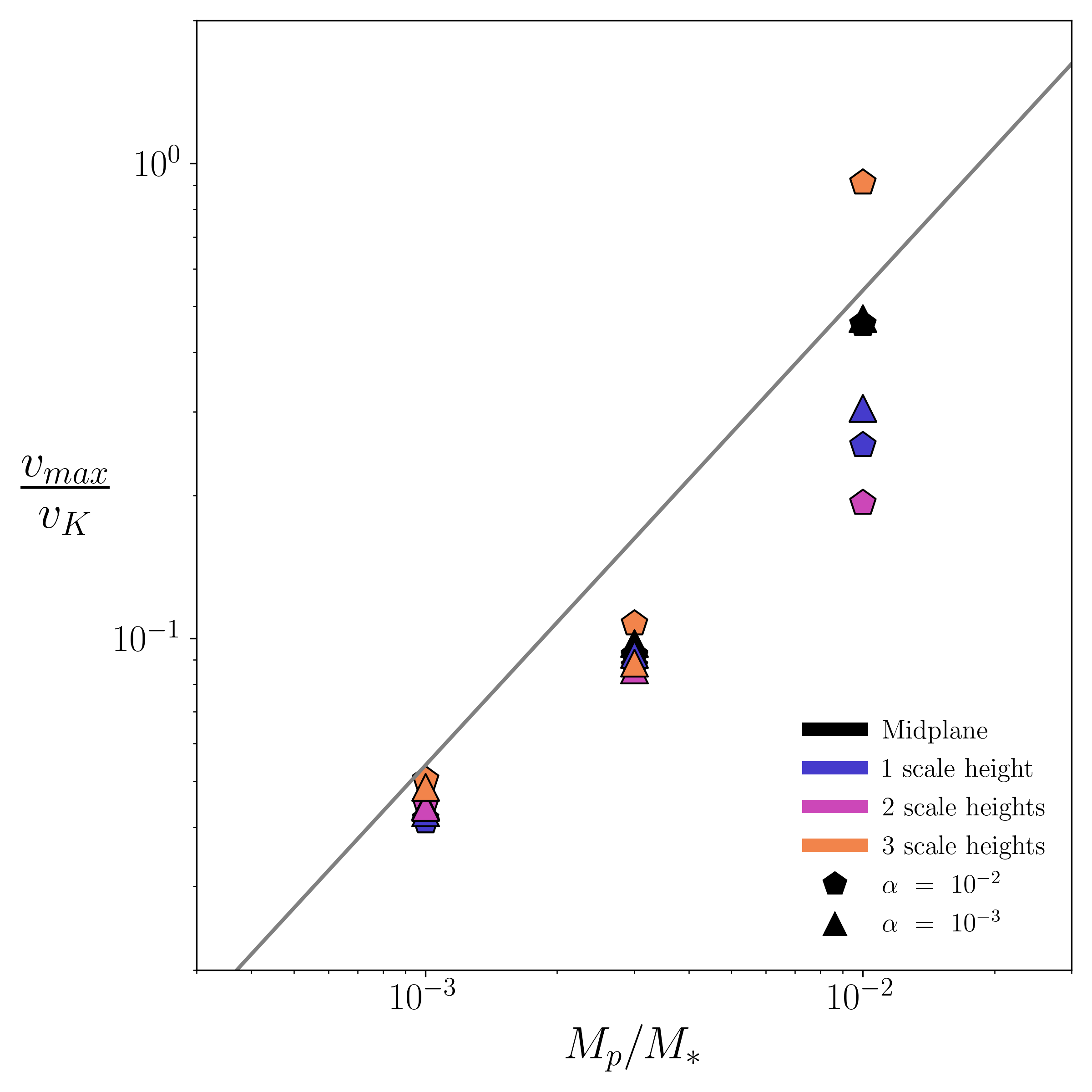}
    \caption{ Maximum deviation from Keplerian velocity $\vel_{max}$ versus planet mass $M_p$, across multiple scale heights of the disk.  $\vel_{max}$ is scaled to the local Keplerian velocity.  Triangle and pentagon markers correspond to disk viscosities of $\alpha=0.001$ and $\alpha = 0.01$, respectively.  The gray line represents a linear relationship between $\vel_{max}$ and $M_p$, as described in the text.}
    \label{fig:vmax}
\end{figure}

After presenting the results on the disk structure without planets, we will study how the planet can influence the disk structure.

Fig. \ref{fig:surfacedensity} shows the disk surface density profiles at $t = 500\ T_0$ when the gap is fully opened by the planet.  Blue curves represent $1M_J$ simulations while orange curves represent $10M_J$ simulations.  The gray shaded regions denote the width of the planet's Hill sphere for 1 and 10 Jupiter masses.  Larger gaps are formed in the presence of higher-mass planets or lower viscosity disks (e.g. \citealt{Fung2014, Kanagawa2015}).

{ Figures \ref{fig:horseshoe} and \ref{fig:horseshoe10MJ} show the velocity components of the disk for the $\alpha=0.01,\ 1 M_J$ and $\alpha=0.01,\ 10 M_J$ simulations, respectively.  Each row shows the disk velocities for a different scale height, with panels showing the vertical velocity $\vel_z$, the radial velocity $\vel_r$, and the deviation from Keplerian velocity $\delta \vel_\phi$. The density panel (at top left) shows the gap and spiral wake.  The streamline of a horseshoe trajectory (starting at the red point) outlines the gap region.  To examine the velocity components across different disk scale heights, we cut the disk at equally spaced values of constant $\theta$.  All of our poloidal cuts are taken on the upper half of the disk; any cuts on the lower half would be almost identical, save for a change in the sign of $\vel_z$ or $\vel_\theta$.  We note that this is a simple way to analyze quantities at different disk scale heights and may not reflect the velocity structure at constant $h/r$ throughout the disk.

The vertical velocity $\vel_z$ (left column) shows the presence of inflows in the vicinity of the planet \citep{Morbi2014,Fung2015}.  The radial velocity (middle column) clearly follows the spirals.  For both the inner and outer spirals at the midplane, the disk region outside the spiral has negative $\vel_r$ while the region at  $r$  smaller than the spiral front has positive $\vel_r$.  The gap is not apparent in the $\vel_r$ panels.  On the other hand, regions of sub/super-Keplerian velocity, shown as $\delta \vel_{\phi}$ (right column), trace the edges of the gap very well. The disk is super-Keplerian (positive $\delta \vel_{\phi}$) at the outer gap edge, while it is sub-Keplerian (negative $\delta \vel_{\phi}$) at the inner gap edge.  This is much clearer in Figure \ref{fig:horseshoe10MJ}, as the velocities in the $1M_J$ simulation are dominated by the overall sub-Keplerian rotation of the gas.  Meanwhile, the spirals are not as visible in the $\delta \vel_{\phi}$ panel. Overall, we conclude that $\vel_r$ traces the spirals while $\delta \vel_{\phi}$ traces the gap.  $\vel_z$ has a smaller amplitude than either $\vel_r$ or $\delta \vel_{\phi}$.  Since our temperature profile is nearly isothermal, we do not observe additional spirals driven by vertical buoyancy resonances \citep{Bae2021}.  However, when present these spirals will appear in the disk velocities as additional spiral signatures traced by the $\vel_r$ and $\vel_z$ components.

We also observe some height dependence for all three velocity components, especially for the $10M_J$ case which has a deep gap.  Around the planet, $\vel_z$ is more negative at 1 scale height than 2 scale heights, implying a higher inflow velocity closer to the planet, while the amplitude of $\delta \vel_{\phi}$ is higher at the midplane decreasing towards the surface.  In the $10M_J$ case, $\vel_{r}$ changes quite dramatically at different disk heights. The large $\vel_r$ around the planet which traces the spirals at the midplane disappears at 1 scale height and only slightly comes back at 2 scale heights, and the $\vel_r$ tracing the spirals at the very inner disk becomes stronger at 2 scale heights. Thus, for massive planets in a deep gap, the kinematic signatures of the planet may show some differences by using various molecular tracers probing different depths in the disk.
}

The region around the planet is significantly affected by the planet in both $\vel_r$ and $\delta \vel_{\phi}$ panels, which can be probed by the distortion in the velocity channel maps \citep{Pinte2018}. The circumplanetary region can be divided into two regions: the region that is bound to the planet (e.g. the circumplanetary disk) and the unbound region which is circulating between the planet and the star \citep{Lubow1999}.  The bound circumplanetary disk has a circular motion around the planet, { and for our simulations it is not completely resolved since the circumplanetary disk is quite small} ($\lesssim$0.4 $R_H$ where $R_H=(M_p/3M_*)^{1/3}a$, \citealt{MartinLubow2011}). Thus, we will only focus on the unbound region for studying the velocity distortion. 

The large velocity in the unbound region is associated with the spirals which are strongest close to the planet.  We calculate the deviation from Keplerian velocity as $\delta \vel_{K} = \sqrt{\vel_r^2 + \delta \vel_\phi^2}$ around the planet. { In Figure \ref{fig:vmax}, we plot $\vel_{max} = \max(\delta \vel_{K})$ for each simulation against the final planet mass $M_p$ for several different scale heights in the disk. This maximum velocity deviation should be close to the maximum ``kink'' velocity measured in observations (e.g. Figure 1 of \citealt{Pinte2018}).  For this Figure, we plot the $1M_J$ and $3M_J$ simulations with a smoothing length of 2 grid cells, as well as the $10M_J$ simulations with the smoothing length scaled to the planet mass.}  To ensure we do not measure the circumplanetary disk, we ignore values within the planet's Hill sphere while calculating $\vel_{max}$. { We tried excluding different sized regions (e.g. 0.4 $R_H$) or using simulations with different smoothing lengths, and found the results to be almost identical.  Above 3 scale heights, $\vel_{max}$ increases dramatically and no longer follows the relationship found at lower scale heights. We believe that these large velocities are numerical in nature, created by the gas density dropping to the simulation density floor in the disk atmosphere. Thus, we did not plot the values beyond 3 scale heights in Figure \ref{fig:vmax}. For the case with the 10 $M_J$ planet in an $\alpha=0.001$ disk, we did not plot $\vel_{max}$ at 2 scale heights and above  where the density reaches the floor inside the deep gap. 

Figure \ref{fig:vmax} shows that, for 1 and 3 $M_J$ cases, $\vel_{max}$ does not depend on $\alpha$ or the disk height where it is measured. 
But for 10 $M_J$ cases, $\vel_{max}$ decreases from the midplane to 2 scale heights and then increases beyond 2 scale heights. This is also apparent from Figure \ref{fig:horseshoe10MJ} where $\vel_r$
becomes small at 1 and 2 scale heights.}

{ Our best fit for the values taken at the disk midplane is plotted as the gray line, which is }
\begin{equation}
\vel_{max} = 54 \vel_K\frac{M_p}{M_*}\,,\label{eq:vmax}
\end{equation}
where $\vel_K$ is the planet's Keplerian velocity around the star. If we apply Equation \ref{eq:vmax} to the observation in \cite{Pinte2018}, the 0.15 $\vel_K$ deviation in their Figure 1 corresponds to  a 5.4 $M_J$ planet around a 1.9 $M_\odot$ star. This is higher than the 2 $M_J$ planet that \cite{Pinte2018} derived by comparing channel maps between observations and direct simulations. Although this difference may be attributed to that we didn't carry out synthetic channel maps, we notice that our $\delta \vel_K$ around the planet ($M_p/M_*=0.001$ cases) is indeed lower than that shown in \cite{Pinte2018} (their Figure 4) by a factor of 2. On the other hand, our values seem to be more consistent with \cite{Perez2018} (their Figure 1). Thus, more detailed comparisons among these simulations with the same parameters are needed in future. { We note that at the disk midplane, our fit in Figure \ref{fig:vmax} has little dependence on the choice of $\alpha$. Even if $\alpha$ changes by one order of magnitude ($\alpha=10^{-3}$ and $10^{-2}$), the maximum velocities of $\delta \vel_K$ for both 1 $M_J$ and 10 $M_J$ cases are almost identical. Thus, we can safely use this relationship to derive the embedded planet mass, with a minimum impact from the unknown disk viscosity.  From this perspective, using the ``kink velocity'' serves as a more robust method to derive the planet mass than using properties of the gap (e.g. \citealt{Zhang2018}) which depends on both $\alpha$ and the gas scale height.} 

The independence of $\vel_{max}$ on $\alpha$ indicates that the maximum velocity of $\delta \vel_K$ is not related to the disk's accretion process. Instead, it is more likely determined by the planet-disk gravitational interaction. { The linear dependence between $\vel_{max}$ and $M_p$ seems to be consistent with the linear theory \citep{Goldreich1979} which shows that, for small amplitude perturbations, the velocity perturbation is proportional to the planet mass. However, the spirals can quickly become highly non-linear when the planet mass is large \citep{Goodman2001,Muto2010,Dong2011,Zhu2013}. The spirals steepen to shocks when they propagate away from the planet at a distance of
\begin{equation}
    |\Delta r_{sh}|\approx 0.93 \left(\frac{\gamma+1}{12/5}\frac{GM_p\Omega_p}{c_s^3}\right)^{-2/5}H\,,
\end{equation}
where $\gamma$ is the adiabatic index. Thus, the shocking distance is $\sim H$ and $\sim 0.4 H$ for the 1 $M_J$ and 10 $M_J$ planet in our $H/R=0.1$ disk. 
Since we only measure $\vel_{max}$ in the region beyond the planet's Hill radius (which is 0.7 $H$ and 1.5 $H$ for 1 $M_J$ and 10 $M_J$ cases), the measured spirals are in the non-linear regime (especially for 10 $M_J$ cases). Furthermore, a deep gap is induced by the massive planet, which clearly indicates that the linear theory cannot be applied to the massive planet case. Thus, we need to sort a different explanation for the measured $\vel_{max}$ with a massive planet.}
For massive planets (e.g. 10 $M_J$), the flow may be quite dynamic. The circumplanetary region is well separated from the background disk, and we expect that the maximum velocity of disk material in a horseshoe orbit  should scale with the free fall velocity to the circumplanetary disk edge. Since the circumplanetary disk has a size of $\lesssim$ 0.4 $R_H$, the unbound flow will have a maximum velocity at the closest approach at $\sim$ 0.4 $R_H$, similar to the escape velocity at $\sim$ 0.4 $R_H$, which is $\vel_{esc}=2.7 \vel_{K} (M_p/M_*)^{1/3}$.
For a 10 $M_{J}$ planet around a solar mass star, $\vel_{esc}=0.58 \vel_K$, which is close to our measurement of 0.5 $\vel_K$. On the other hand, the escape velocity scales with $M_p^{1/3}$, while our fitting shows a linear scaling with $M_p$. For a $M_J$ planet, the escape velocity at 0.4$R_H$ is higher than our measured $\vel_{max}$. We attribute this to the fact that the gas flow is significantly affected by the gas pressure within the shallow gap induced by a low mass planet and the linear perturbation theory may apply.


\begin{figure*}
 \includegraphics[width=0.98\textwidth]{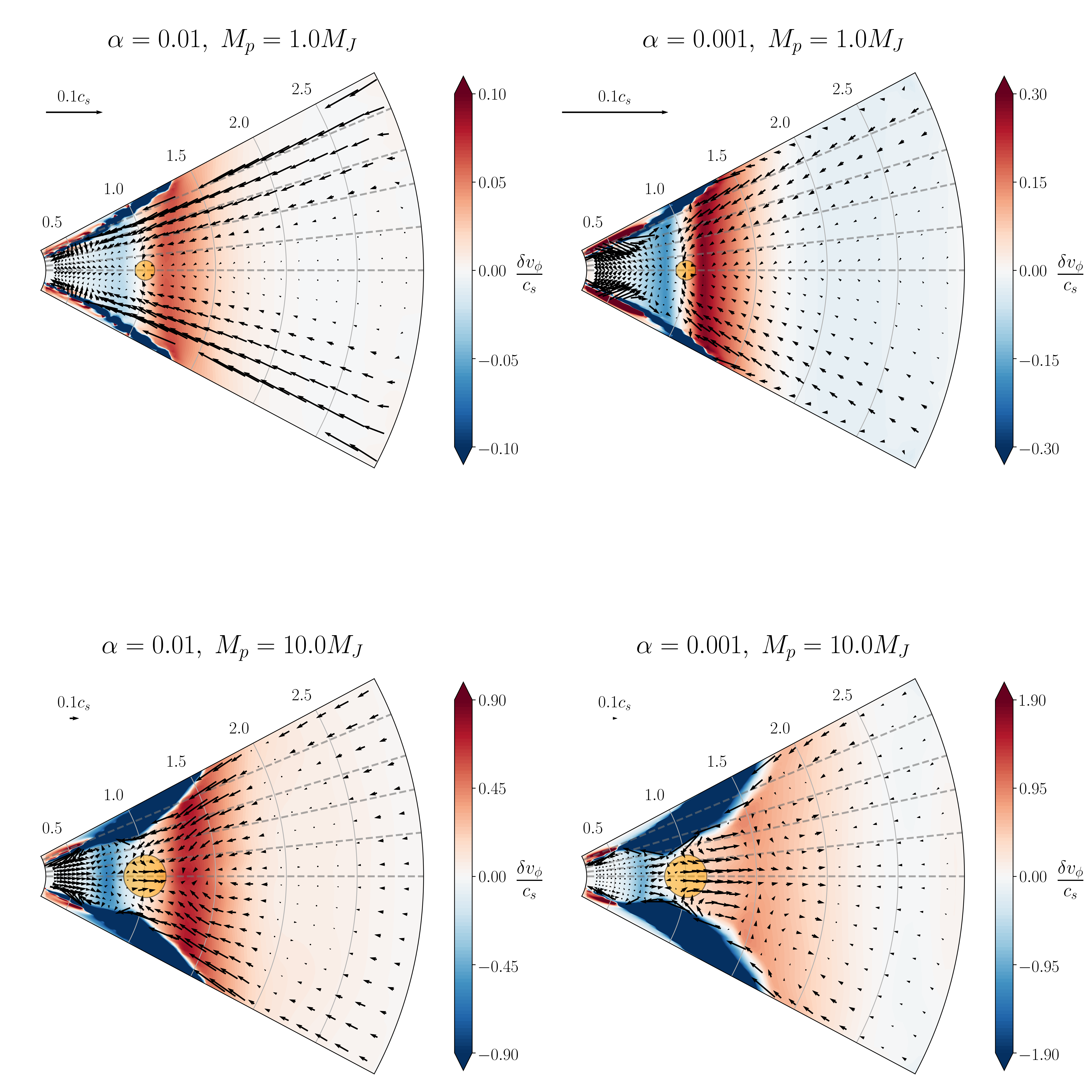}
 \caption{Azimuthally averaged velocity structure for each of the 3D simulations.  Black arrows denote the velocity flow in the $(r, \theta)$ plane.  Background colors denote the deviation from normal Keplerian rotational velocity, with red colors indicating super-Keplerian flow and blue colors indicating sub-Keplerian flow.  The orange circle denotes the planet's Hill radius.  Gray dashed lines mark the disk scale heights.  Some vectors are omitted for clarity; see text for details.}
 \label{fig:disk}
\end{figure*}

\begin{figure*}
 \includegraphics[width=\textwidth]{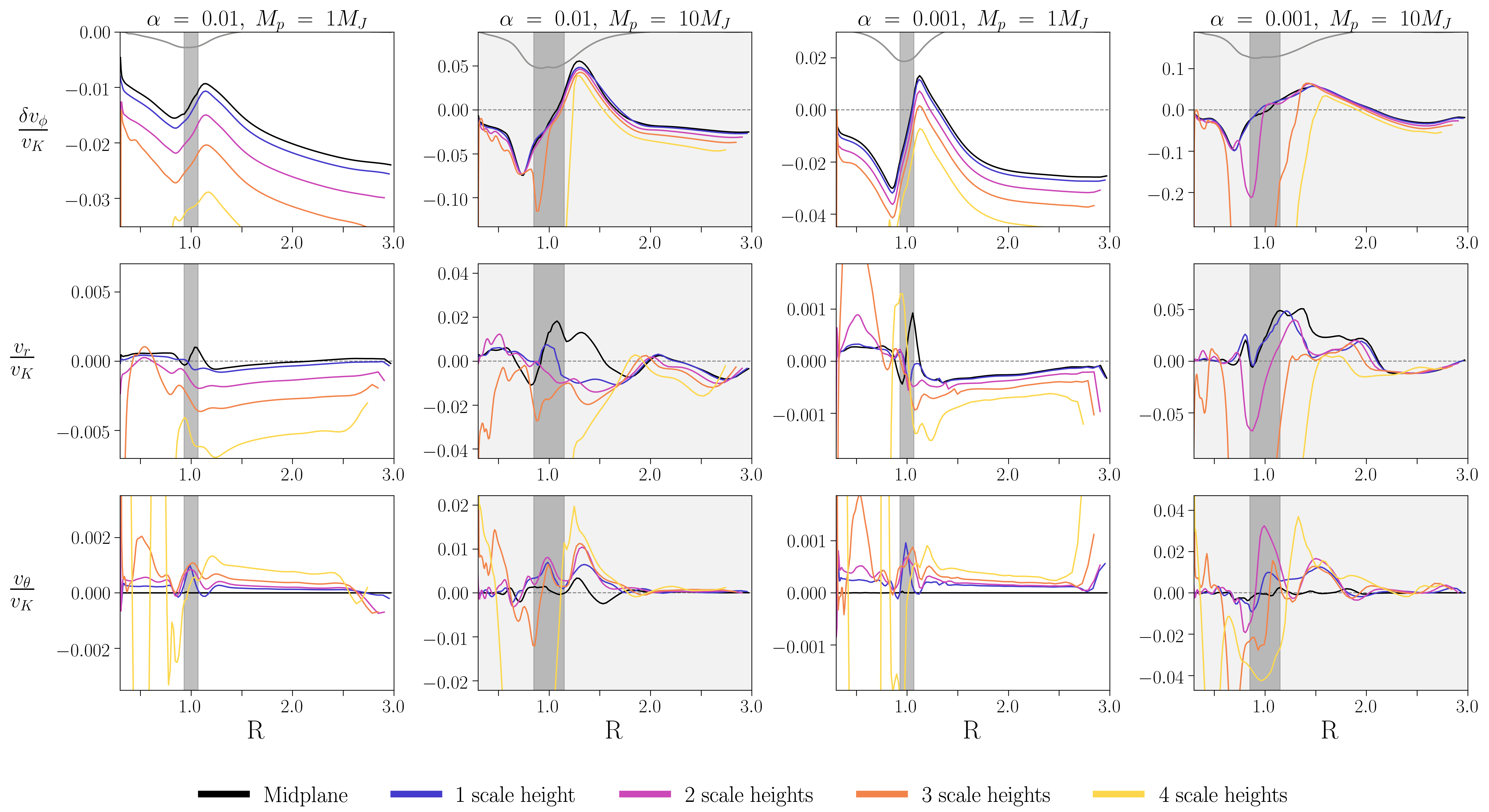}
 \caption{Azimuthally averaged velocity profiles for the 3D simulations.  Each row plots a velocity component versus \textit{cylindrical} R across multiple scale heights of the disk.  \textit{Top Row}: differential azimuthal velocity. Gray curves are surface density curves shown in Fig. \ref{fig:surfacedensity}. \textit{Middle Row}: Radial velocity.  \textit{Bottom Row}: Poloidal velocity.   All speeds are normalized to the local Keplerian speed.  Different colors represent different scale heights in the disk.  Shaded regions represent the extent of the planet's Hill radius.}
 \label{fig:velprofile}
\end{figure*}

\subsection{Azimuthally-averaged Velocity Structure Induced by the Planet}
\label{sec:sumplot}

\begin{figure*}
 \includegraphics[width=\textwidth]{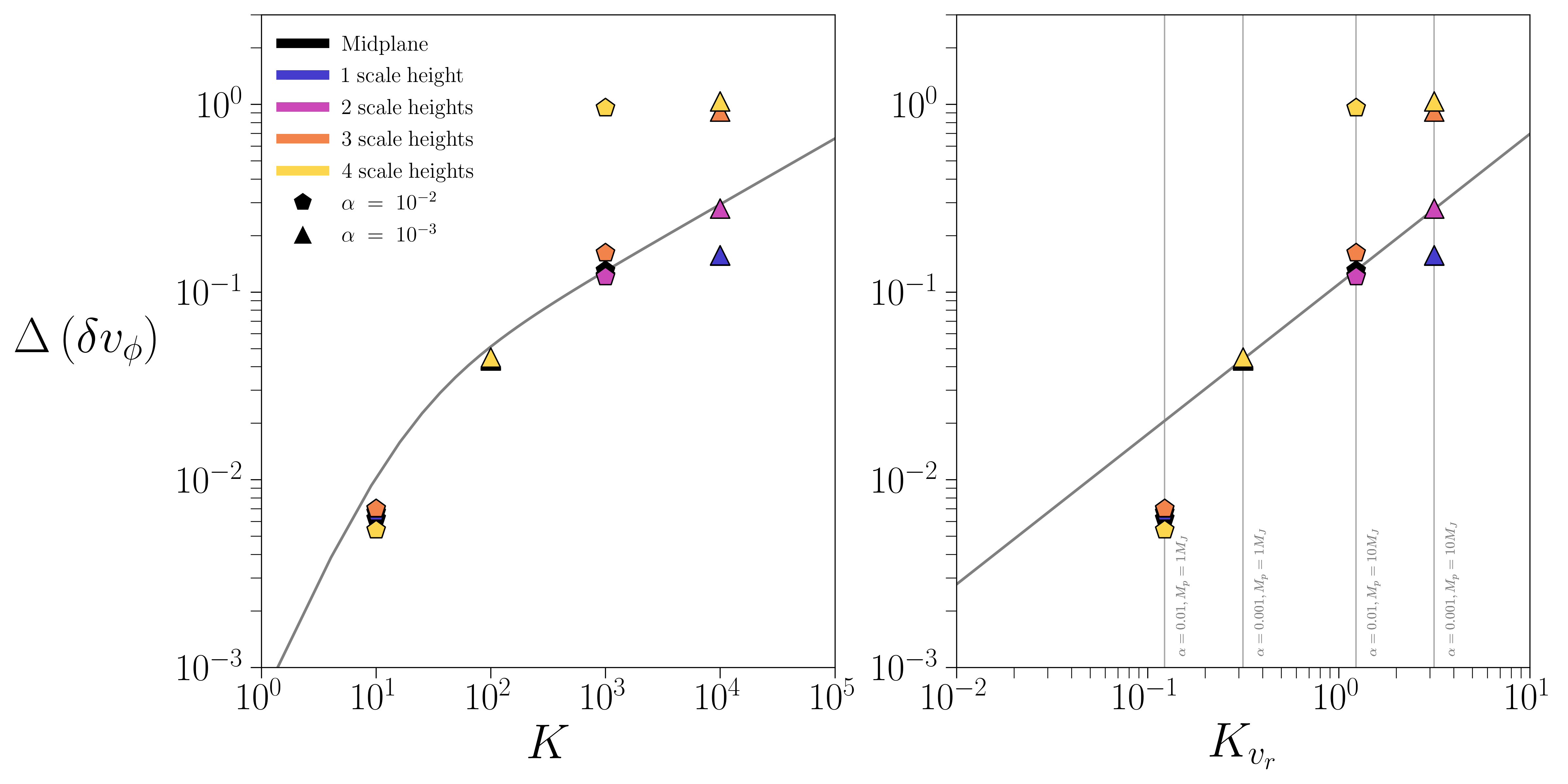}
 \caption{The deviation amplitude $\Delta (\delta \vel_{rot})$ plotted for different fitting parameters $K$.  \textit{Left}:  Using the fitting parameter from \protect\cite{Kanagawa2015} and best-fit equation from \protect\cite{GyeolYun2019}.  Right:  Using the fitting parameter and best-fit equations from \protect\cite{Zhang2018}.  Best-fit equations are shown in gray.  Colors are used to represent different scale heights, as in Fig. \ref{fig:velprofile}.  The triangle and pentagon markers correspond to disk $\alpha$-viscosities of $10^{-3}$ and $10^{-2}$, respectively, following the convention used in \protect\cite{Zhang2018}. }
 \label{fig:summary}
\end{figure*}

Figure \ref{fig:disk} shows the velocity structure of the disk, azimuthally averaged in the $r-\theta$ plane at $t=500\ T_0$.  All velocities are scaled relative to the local sound speed.  The  poloidal velocity is shown as black arrows, while the background color denotes the gas rotational speed.  Blue and red areas denote regions of sub- and super-Keplerian rotation, respectively.  Meridional flows are visible in all simulations around the planet's Hill radius (denoted by the orange circle at $R=1$).  The circulation is stronger in disks with a deeper gap.  To maintain figure readability, we only display vectors every three grid cells, and exceedingly large velocities (which typically occur towards the disk atmosphere) have been omitted.

Although the meridional flow shown in Figure \ref{fig:disk} is regarded as an evidence of the presence of a planet \citep{Teague2019}, such flow motion is highly non-axisymmetric in nature. As shown in Figure \ref{fig:vslice}, although the planet's gravity leads to strong inflow to the planet from the planet's polar direction \citep{Tanigawa2012, Fung2015, Szulagyi2016}, such inflow motion is localized around the planet. At $\phi=\phi_p+\pi$, there is no such meridional motion present.  Thus, understanding whether the observed meridional motion is truly axisymmetric in disks (e.g. similar meridional patterns at different values of $\phi$) or whether it only occurs in some local disk region is important for confirming its planet origin.  

A detailed examination of the azimuthally-averaged velocity is shown in Fig. \ref{fig:velprofile}.  { We examine each velocity component across the entire (cylindrical) radial extent and vertical extent of the disk, using the same polar cuts used in Figures \ref{fig:horseshoe} and \ref{fig:horseshoe10MJ} to examine different scale heights.  The disk scale heights are visible in Figure \ref{fig:disk} as gray dashed lines in the upper half of the disk. }  
We plot the azimuthally averaged velocities at the end of the simulation ($t=500\ T_0$), and remove all cells within the Hill sphere of the planet to remove local effects from the planet's vicinity.  Since the number of $\theta$ cells in our simulations is even, the midplane velocity is calculated by averaging the two cells just above and below the midplane.

The $\delta \vel_{\phi}$ panels (top panels) in Figure \ref{fig:velprofile} show a vertical shift as one moves vertically upward from the disk midplane, a feature we attribute to the gas pressure gradient.  The displacements are more noticeable in disks with smaller perturbations relative to the disk Keplerian speed (this is also true relative to the local sound speed $c_s$, since $h/r = c_s/\vel_\phi$ is assumed a constant vertically).  The magnitude of the velocity shift induced by the pressure gradient is at most a few percent of the Keplerian velocity.  In spite of the vertical shift, the shape of the curve is unchanged below 3 scale heights.  By comparing these curves with the surface density plots (redrawn at the top of the figure), we see that the points of minimum and maximum deviation correspond to the points of steepest pressure gradient, and the magnitude of the deviation is roughly equal to the value of the background pressure gradient at the gap edges, similar to that described in \cite{Teague2018}. Above 3 scale heights, the residual velocities in the inner disk become highly negative, larger than expected for the given planet and disk parameters. 

\cite{Zhang2018} and \cite{GyeolYun2019} have studied the deviation from Keplerian velocity produced at the gap edge and how the velocities change with different disk parameters. On the other hand, these works are based on 2-D simulations, so that they focus on the velocity deviation at the disk midplane, while most ALMA molecular line observations actually probe the disk surface \citep{Teague2018, Teague2019}. These studies have found relationships between the parameters of the planet-disk system and the amplitude of the velocity deviations $\dvrot$ (measured from the sub-Keplerian minimum interior to $R_p$ to the super-Keplerian maximum exterior to $R_p$).  The dependence of the planet-disk parameters are commonly expressed in terms of a fitting parameter $K$, where $K = q^{pa} \left( \frac{h}{r} \right)^{pb} \alpha^{pc}$ and the power-law exponents $pa$, $pb$, and $pc$ are to be determined.

\cite{GyeolYun2019} run a parameter study using the fitting parameter $K = q^{2} \left( \frac{h}{r} \right)^{-5} \alpha^{-1}$ from \cite{Kanagawa2015}.  They find a fit of

\begin{equation}
\label{eq:dvgyeolyun}
    \dvrot = \left( \frac{h}{r} \right) \frac{0.007 K^{1.38}}{1 + 0.06K^{1.03}}
\end{equation}

In a different parameter study by \cite{Zhang2018},  both $K$ and $\dvrot$ are fit to the simulation data, with:

\begin{equation}
    K_{\vel_r} = q \left( \frac{h}{r} \right)^{-1.27} \alpha^{-0.41} \,,
\end{equation}
and
\begin{equation}
\label{eq:dvzhang}
    \dvrot = 0.11 K_{\vel_r}^{0.80} \,.
\end{equation}

Both of these parameter studies used 2D hydrodynamic simulations, and thus were unable to study how $\dvrot$ changes at different disk scale heights.  In Figure \ref{fig:summary}, we plot $\dvrot$ at different disk heights with respect to $K$ and $K_{\vel_r}$ in our simulations, together with the two fitting formulas (Equation \ref{eq:dvgyeolyun} on the left and Equation \ref{eq:dvzhang} on the right).  We can see that both formulas fit equally well at large $K$ and $K_{\vel_r}$ values. But Equation \ref{eq:dvgyeolyun}  fits slightly better than Equation \ref{eq:dvzhang} for small $K$ and $K_{\vel_r}$ (which means smaller mass planets). Most importantly, $\dvrot$ is almost a constant within 3 disk scale heights. For larger values of $K$ and $K_{\vel_r}$ (e.g. disks with larger planets, smaller scale heights, or lower values of $\alpha$), the value of $\dvrot$ begins to deviate above 3 scale heights. { This indicates that the deviation from the Keplerian velocity increases towards the disk surface faster in a deeper gap. }

At 4 scale heights, closer to the disk atmosphere, we find that $\dvrot$ departs strongly from the expected behavior and increases substantially to roughly order unity.  Since $\delta \vel_\phi = \vel_\phi - R \Omega_K = \vel_\phi-\vel_K$, the large values of $\dvrot$ suggests that $\delta \vel_\phi$ is changing by nearly the local Keplerian velocity in the vicinity of the gap.  Examining the azimuthal velocities in Figure \ref{fig:velprofile} shows that the super-Keplerian peak in $\delta \vel_\phi$ is $\lesssim 10\%$ of Keplerian velocity at 4 scale heights, suggesting that most of the contribution to $\dvrot$ is from sub-Keplerian motion.  Thus, a possible interpretation of this feature is that, in the vicinity of the gap, material in the disk atmosphere has lost nearly all of its rotational velocity and is instead falling directly towards the star.

Overall, as long as the molecular tracers \citep{Isella2018}  are tracing the disk region within $\sim$ 3 disk scale heights, we can use Equations \ref{eq:dvgyeolyun} and \ref{eq:dvzhang} to derive the planet mass from the change of the azimuthal velocities. If the tracers are tracing the region beyond $\sim$ 3 disk scale heights, using Equations \ref{eq:dvgyeolyun} and \ref{eq:dvzhang} can overestimate the planet mass.  { Emission from different molecular tracers is visible at different vertical disk heights due to different optical depths \citep{Flaherty2017}, so comparing observations from different molecular tracers, such as the different CO isotopologues, may allow observations to distinguish velocities at different disk scale heights. }

The disk radial velocities (Fig. \ref{fig:velprofile}, middle row) show a similar profile to the azimuthal velocity in the vicinity of the planet.  These features are the signature of the "planet-driven" flows where the planet moves material out of the gap.  Interior to $r_p$, there is a decrease in the radial velocities (even to negative values), while exterior to $r_p$ there is an increase.  This is most apparent in the $M_p = 10 M_J$ simulations.  These flows are strongest at the disk midplane and weaken as one moves vertically in the disk; for most of our simulations, these flows are no longer visible above 2 scale heights.  Some simulations show a reversed radial flow direction in the disk atmosphere, with $\vel_r$ positive interior to $r_p$ and negative exterior to $r_p$.  This reversal may correspond to the material at the surface flowing into the gap.  The appearance of these local extrema in the radial velocity channel is a signature unique to the planet and may be important in determining the existence of a planet in an observed gap.  Section \ref{sec:discussion} examines this idea further by comparing the velocity fields of planetary and non-planetary gaps.

Gas inflows are also visible in the disk poloidal velocities (Fig. \ref{fig:velprofile}, bottom row), as velocity peaks at the planet's orbital radius (except for the velocity at 4 scale heights).  Since the data in Figure \ref{fig:velprofile} are taken from the upper half of the disk, the inflows appear as a positive bump near $r=1$. The inflow pattern is visible for several scale heights, though the magnitude of these features is not as large as the $\delta \vel_\phi$ signature.   We note that, for the $10 M_J$ simulations, additional downward motion at $R\sim1.5$ appear outside of the planet's orbital radius; this is also visible in Figure \ref{fig:disk}.  
\section{Discussion}
\label{sec:discussion}

\subsection{Predicting The `Kink Velocity' For The DSHARP Sample}
\begin{table*}
    \centering
    \begin{tabular*}{0.9\textwidth}{p{0.10\textwidth}
                                 p{0.05\textwidth}
                                 p{0.10\textwidth}
                                 p{0.05\textwidth}
                                 p{0.05\textwidth}
                                 p{0.10\textwidth}
                                 p{0.05\textwidth}
                                 p{0.05\textwidth}
                                 p{0.08\textwidth}
                                 p{0.05\textwidth}}
        \hline
        Disk & Gap (AU) & $M_* (M_\odot)$ & $M_{p, min}$ & $M_{p}$ & $M_{p, max}$ & $\delta \vel_{min}$ & $\delta \vel$ & $\delta \vel_{max}$ & $\delta \vel_{obs}$\\
        \hline
        AS 209      & 9     & 0.83 & 0.37  & 3.38  & 4.18  & 0.0241   & 0.22     & 0.272    & -   \\
        AS 209      & 99    & 0.83 & 0.18  & 0.75  & 1.32  & 0.0117   & 0.0488   & 0.0859   & - \\
        Elias 24    & 57    & 0.78 & 0.19  & 0.81  & 1.72  & 0.0132   & 0.0561   & 0.119    & -  \\
        Elias 27    & 69    & 0.49 & 0.02  & 0.1   & 0.12  & 0.0022   & 0.011    & 0.0132   & ? \\
        GW Lup      & 74    & 0.46 & 0.01  & 0.06  & 0.06  & 0.00117  & 0.00704  & 0.00704  & < 0.3 \\
        HD 142666   & 16    & 1.58 & 0.09  & 0.5   & 0.62  & 0.00308  & 0.0171   & 0.0212   & - \\
        HD 143006   & 22    & 1.78 & 2.35  & 9.81  & 40.6  & 0.0713   & 0.298    & 1.23     & $\approx$ 0.20 \\
        HD 143006   & 51    & 1.78 & 0.14  & 0.57  & 0.67  & 0.00425  & 0.0173   & 0.0203   & - \\
        HD 163296   & 10    & 2.04 & 0.19  & 1.18  & 1.46  & 0.00503  & 0.0312   & 0.0386   & - \\
        HD 163296   & 48    & 2.04 & 0.54  & 2.24  & 4.45  & 0.0143   & 0.0593   & 0.118    & - \\
        HD 163296   & 86    & 2.04 & 0.08  & 0.34  & 0.34  & 0.00212  & 0.009    & 0.009    & $\approx$ 0.15 \\
        SR 4        & 11    & 0.68 & 0.38  & 3.57  & 4.41  & 0.0302   & 0.283    & 0.35     & - \\
        \hline
    \end{tabular*}
    \caption{ Estimated velocity deviations for selected disks using Equation \ref{eq:vmax}, in units of the local Keplerian velocity.  Distance and stellar mass data from \protect\cite{Andrews2018} and planet mass estimates from \protect\cite{Zhang2018}.  Gap distances are measured in AU and used as the planet-star distance.  Star masses are given in $M_\odot$, and planet masses are given in $M_J$.  The ``middle'' planet mass estimates are taken from the middle values of Column 13 in \protect\cite{Zhang2018}.  Values in $\delta \vel_{obs}$ are the maximum velocity deviations from \protect\cite{Pinte2020}, where applicable.}
    \label{tab:disk_vmax}
\end{table*}

{ The velocity disturbances created by the planet in Figures \ref{fig:horseshoe} and \ref{fig:horseshoe10MJ} trace features that span across the entire disk, some of which are present across multiple scale heights.  These disturbances are strongest in the vicinity of the planet, and so good azimuthal resolution is desirable in observations.  For disks with high azimuthal resolution (roughly $\lesssim 1 R_H$), deviations from Keplerian velocity can be localized to specific areas in the disk, which has been used to infer the presence and position of planets using velocity channel maps (\citealt{Pinte2018, Pinte2020}).  

Given a mass for the embedded planet, Equation \ref{eq:vmax} can be used to predict the velocity deviation created in the disk.  In Table \ref{tab:disk_vmax}, we calculate the expected velocity deviations ('kink velocity') for planets in selected disks.  We use planet masses from \cite{Zhang2018}. For the 22 AU gap in HD 143006, our calculated value for $\delta \vel$ is roughly consistent with the measured value obtained in \cite{Pinte2020}. On the other hand, most planet masses constrained by \cite{Zhang2018} for the DSHARP sample are less than $M_J$, which corresponds to $\vel_{max}/\vel_K\lesssim $0.05 (Figure \ref{fig:vmax}), which requires high spectral-resolution observations. For some cases (e.g. the 86 AU planet candidate in HD 163296), our predicted `kink'
velocity is significantly smaller than observations in \cite{Pinte2020}, which indicates some inconsistency between the planet mass estimate using the dust continuum method and 
the mass estimate using the gas kinematic method. To understand this inconsistency, it is crucial to constrain the gaseous gap depth and width, which can also be used to derive the
planet mass (e.g. Figure \ref{fig:summary})  without assuming some dust size in the disk (as in the dust continuum method).} 

\subsection{Azimuthally-averaged Velocity Structure For A Gap Without A Planet}
{ Although Figure \ref{fig:summary} links the planet mass with the gas kinematics at the gap edge, we want to note that the gas kinematics measurements in \cite{Teague2018} and \cite{Teague2019} are actually signatures of a local decrease in the gas surface density, and they are not direct probes for the embedded planet. } For the gas, there is a radial force due to the pressure gradient $\frac{dP}{dr}$, and the change in the pressure gradient due to the gap causes a change in the azimuthal velocity $\vel_\phi$.  The strength of the meridional flows observed in the disk atmosphere is also a function of gap depth, as the flows angle more strongly in the polar direction over the gap.  However, neither of these is a direct signature of an embedded planet; a gap created by non-planetary means with a similar gap depth would exhibit the same deviations in azimuthal velocity and gap inflows, despite no planet existing within the gap.

Thus, we want to compare the velocity fields in gaps created by planetary and non-planetary methods in order to identify any velocity signatures that are unique to the planet.  To do this, we extend our simulations by removing the planetary mass and simulating an additional time of $50\ T_0$.  Without the gravitational influence of the planet, any unique dynamic planetary signatures are dispersed within several orbits. This roughly mimics the gaps created by non-planetary methods which may operate more axisymmetrically, { such as snow lines and MRI}, even though no additional physics has been implemented in our model. { We note that the gap slowly closes over tens ($\alpha=0.01$ cases) or hundreds ($\alpha=0.001$ cases) of 
orbits due to the viscous spreading.}

\begin{figure}
 \includegraphics[width=\columnwidth]{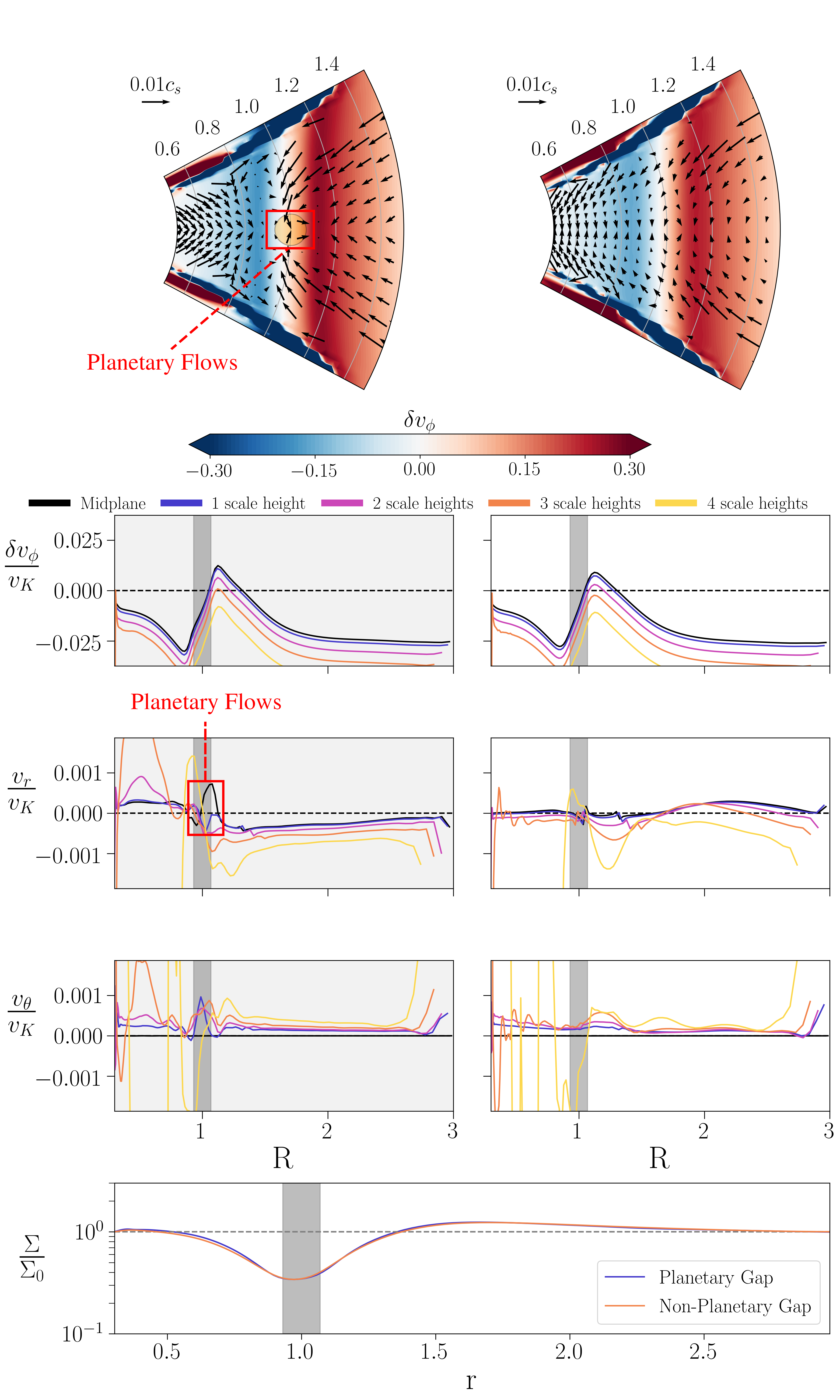}
 \caption{Comparison of the disk velocity for "planetary" and "non-planetary" gaps in the $\alpha = 0.001$,\ $M=1M_J$ simulation.  \emph{Left column:} snapshot with a planet opening the gap. \emph{Right column:} snapshot 10 orbits after removing the planet from the gap.  \emph{From top to bottom:}  Azimuthally averaged velocity profile in the $r-\theta$ plane, zoomed in on $r=1$; Azimuthally averaged azimuthal, radial, and poloidal velocity components, as presented in Fig. \ref{fig:velprofile}; surface density profiles.  The snapshots are chosen so that the gap depths for the disks with and without the planet are nearly identical. }
 \label{fig:gapcomp}
\end{figure}

A snapshot of the original, ``planetary'' gap is compared to the new, ``non-planetary'' gap in Figure \ref{fig:gapcomp} for the simulation with $\alpha = 0.001$ and $M_p = 1M_J$. The ``non-planetary'' gap is chosen at 10 orbits after removing the planet from the gap.  The snapshot of the ``planetary'' gap is chosen such that the gap depth $\Sigma/\Sigma_0$ in both snapshots is nearly the same.  Though we only show one simulation in this figure, the behavior is roughly the same across all of our simulations.  The left column shows both 1D and 2D profiles for the planetary gap, while the right column shows the same for the non-planetary gap.  Comparing the azimuthal velocities, we see that the deviations from normal rotational velocity are nearly identical for both gaps along the full vertical extent of the disk.  Thus, a measurement of $\delta \vel_\phi$ alone is not sufficient evidence to confirm the existence of an embedded planet.

{ Both the $\vel_r$ and $\vel_\theta$ components are also similar between the ``planetary'' gap and the ``non-planetary'' gap with the circulation pattern at the gap edge. The only noticeable difference is that, in the planetary gap, both the $\vel_r$ and $\vel_\theta$ components at the midplane (black curves) show peaks in the vicinity of the planet's orbital radius.  The $\vel_r$ peak represents the repelling radial motion away from the planet which is probably driven by the spirals or the gravitational influence of the planet, and the $\vel_\theta$ peak represents infall onto the planet. Thus these peaks are signatures of the planet itself.}  However, these signatures are roughly an order of magnitude smaller than the $\delta \vel_\phi$ signature, making their detection in velocity maps particularly challenging.  Their vertical extent is also quite limited; the outflows present in the $\vel_r$ component weaken quickly with height (disappearing at one scale height), and so are likely not visible with CO measurements.  The inflows that cause a signature in the $\vel_\theta$ component have a larger vertical extent, of 2-3 scale heights. Another planetary signature which is not captured in our azimuthally averaged approach is the broader line width within the gap due to strong turbulence induced by the planet \citep{Dong2019}.

We also compare our constant-$\alpha$ simulations (with $\alpha=0.001$) to our variable-$\alpha$ simulation, which uses the stress profile described by Equation \ref{eq:unistress}.  Both simulations have $M_p = 1M_J$.  The two models are very similar, with most of the global features remaining the same.  The gap in the variable-$\alpha$ model is somewhat shallower than the constant-$\alpha$ model, by a factor of about 1.4.  The deviation amplitude in $\delta \vel_\phi$ is slightly smaller than the constant-$\alpha$ simulation, and so the measured value of $\dvrot$ would also be smaller.  Close to $z = 4H = h_{cut}H$, the region where the stress profile changes, the gas velocity increases sharply.

\section{Conclusions}
\label{sec:conclusion}
We have studied how ALMA kinematic observations can help us to constrain 1) the accretion mechanisms, and 2) the planet properties.

We have examined how different disk stress profiles affect its velocity structure, and how the velocity structure in turn affects its subsequent evolution.  We find that the radial velocity profile of the disk is very sensitive to the stress profile that is chosen. Thus, future kinematic observations using various molecular lines tracing different disk heights will not only measure the value of $\alpha$ but also constrain the detailed vertical stress profiles and accretion mechanisms. 

On the other hand, as long as the vertically integrated stress is the same, the evolution of the disk's global surface density is unaffected.  We also find that steep dropoffs at the outer edge of the disk, which are normally explained by phenomenon external to the disk such as truncation or photoevaporation, can also be explained by a disk with a radially-varying $\alpha$ profile.

In our study of three-dimensional velocity flows of a planet opening a gap in the disk, we are able to see how different velocity components follow different components of the planet-disk system.  The radial velocity $\vel_r$ traces out the spiral wake of the planet, while the sub-/super-Keplerian velocity $\delta \vel_\phi$ traces out the edges of the gap. We observe some dependence of these features on the disk height, especially for massive planet cases which have deep gaps.
The linear relationship between the planet mass and the ``kink velocity'' is derived, and it is independent from the disk viscosity and the disk height (except for very massive planet cases). Using such a relationship, we predict the ``kink velocity'' for the planet candidates in the DSHARP sample.

The velocity deviation at the gap edge within 3 disk scale heights is consistent with previous 2D studies at the midplane. We see meridional circulation in the azimuthally averaged velocity maps, and are able to identify components of the circulation process across the vertical extent of the disk.  However, by comparing the gap carved by the planet to a gap that is non-planetary in nature, we find that the deviation from Keplerian rotation and the meridional circulation is a feature of the gap and not necessarily a signature of the planet itself.  Examining the velocity field throughout the disk reveals some velocity signatures that are unique to the planet, but they are highly non-axisymmetric, limited in their vertical extent, and are an order of magnitude smaller than the deviations caused at the gap edge.  Thus, they are not easily detectable in azimuthally averaged velocity maps and may also be difficult to detect with current observations. 
{ 
Combining both axisymmetric kinematic observations and the residual “kink” velocity is needed to probe young planets in protoplanetary disks. More specifically, we can derive the gaseous gap depth and width using the planet mass constrained by the ``kink'' velocity, and then study if the predicted velocity structure at the gap edge are consistent with axisymmetric kinematic observations.
}

\section*{Acknowledgements}

All simulations are carried out using computers supported by the Texas Advanced Computing Center (TACC) at the University of Texas at Austin through XSEDE grant TG-AST130002 and from the NASA High-End Computing (HEC) program through the NASA Advanced Supercomputing (NAS) Division at Ames Research Center.  Z. Z. acknowledges support from the National Science Foundation under CAREER grant AST-1753168.

\section*{Data Availability}
The data used in this paper is available upon request to the corresponding author.




\bibliographystyle{mnras}
\bibliography{ref} 







\bsp	
\label{lastpage}
\end{document}